\documentclass[%
 aps,
 amsmath,amssymb,
 reprint,%
]{revtex4-1}

\def\ket#1{|\,#1\,\rangle}
\def\bra#1{\langle\, #1\,|}

\def\opone{\leavevmode\hbox{\small1\kern-3.8pt\normalsize1}}
\newcommand{\fla}[1]{\begin{flalign}#1\end{flalign}}
\def\abs#1{|\,#1\,|}

\usepackage{color}
\usepackage{graphicx}
\usepackage{dcolumn}
\usepackage{bm}

\usepackage[utf8]{inputenc}
\usepackage[T1]{fontenc}
\usepackage{mathptmx}
\usepackage{etoolbox}

\makeatletter
\def\@email#1#2{%
 \endgroup
 \patchcmd{\titleblock@produce}
  {\frontmatter@RRAPformat}
  {\frontmatter@RRAPformat{\produce@RRAP{*#1\href{mailto:#2}{#2}}}\frontmatter@RRAPformat}
  {}{}
}%
\makeatother
\begin{document}

\title{Photonic interface between subcarrier wave and dual rail encodings}
\author{K. S. Melnik}
\affiliation{ 
Kazan Quantum Center, Kazan National Research Technical University, 18a Chetaeva str.,  Kazan, 420111, Russia}%
\author{E. S. Moiseev}%
\affiliation{ 
Kazan Quantum Center, Kazan National Research Technical University, 18a Chetaeva str.,  Kazan, 420111, Russia}%
 \email{e.s.moiseev@kazanqc.org}

\begin{abstract}
Quantum key distribution with multimode subcarrier wave encoding is propitious for being robust against environmental disturbance.
For application in long-distance quantum communications, this encoding has to be compatible with entanglement-assisted quantum repeaters that are commonly designed to work with dual rail encodings.  
We propose and demonstrate an interface between subcarrier wave and dual rail encodings with a fidelity of $0.95 \pm 0.03$ using a linear optical circuit.
The developed scheme opens the way for future heterogeneous quantum networks that combine subcarrier wave quantum key distribution with trusted and non-trusted nodes.
\end{abstract}

\maketitle
\section{Introduction}
The communication distance of point-to-point quantum key distribution (QKD) is fundamentally limited by losses in the optical channel to hundreds of kilometres \cite{Pirandola2017}. 
Current long-distance QKD systems use a chain of connected trusted nodes that translate the generated key from one end of the chain to another \cite{Elliott_2002, Sasaki:11,Dynes2019,Chen2021}. 
Different from point-to-point QKD protocols, the twin-field QKD improves the secure key rate scaling with distance. The recent result demonstrates long distance QKD and may provide a solution to reduce the number of trusted relays \cite{Xu20,Chen20}.
    Alternatively, the development of the quantum internet pursuits the creation of entanglement-enriched quantum networks, such as quantum repeater, that could be used for QKD without a need of the trusted nodes \cite{Briegel98, Guha2015}. For a cohesive operation of the existing QKD lines with the future quantum networks, a mutual encoding of quantum information is necessary.
Conventional discrete variable quantum repeater provides photonic entanglement in the form of the Bell pairs in dual rail encoding (e.g., polarization) \cite{Sangouard11,Yu2020}.

The creation of interfaces between different quantum informational encodings is a significant milestone for the development of a quantum internet \cite{Wehner2018}. It would allow for the connection of local quantum networks with different encodings and their combination into a global heterogeneous network with tolerable losses at the interconnections.
Recently, several experiments were performed on conversion between dual and single rail qubits \cite{drahi2021Hybrid}, teleportation-based conversion between single rail  and wave-like qubits \cite{Ulanov2017}, between dual rail and wave-like qubits \cite{Sychev2018} and on the interface between hybrid and dual rail entanglement \cite{Guccione20} to name a few.

The subcarrier wave (SCW) encoding is a promising for point-to-point \cite{PhysRevLett.82.1656,Gleim:16}, Plug\&Play \cite{Bannik:21}, continuous variables \cite{Melnik2018,SCW-CV} and twin-field QKD \cite{Chistiakov:19} due to its robustness to environmental effects acting on a fiber line, interferometer-free optical scheme and its capacity for spectral multiplexing \cite{mora2012simultaneous}.
In SCW-QKD the information is encoded in a value of the phase of the applied sinusoidal optical phase modulation, that leads to specific multi-mode state.
However, this multi-mode state complicates a creation entanglement distribution protocol for this encoding that limits potential long-distance SCW quantum communications.
One way to circumvent this issue would be a conversion of an information from SCW encoding into the one with available quantum repeater protocol, e.g. dual rail encoding.

\begin{figure*}[ht]
\centering
\includegraphics[width=0.95\linewidth]{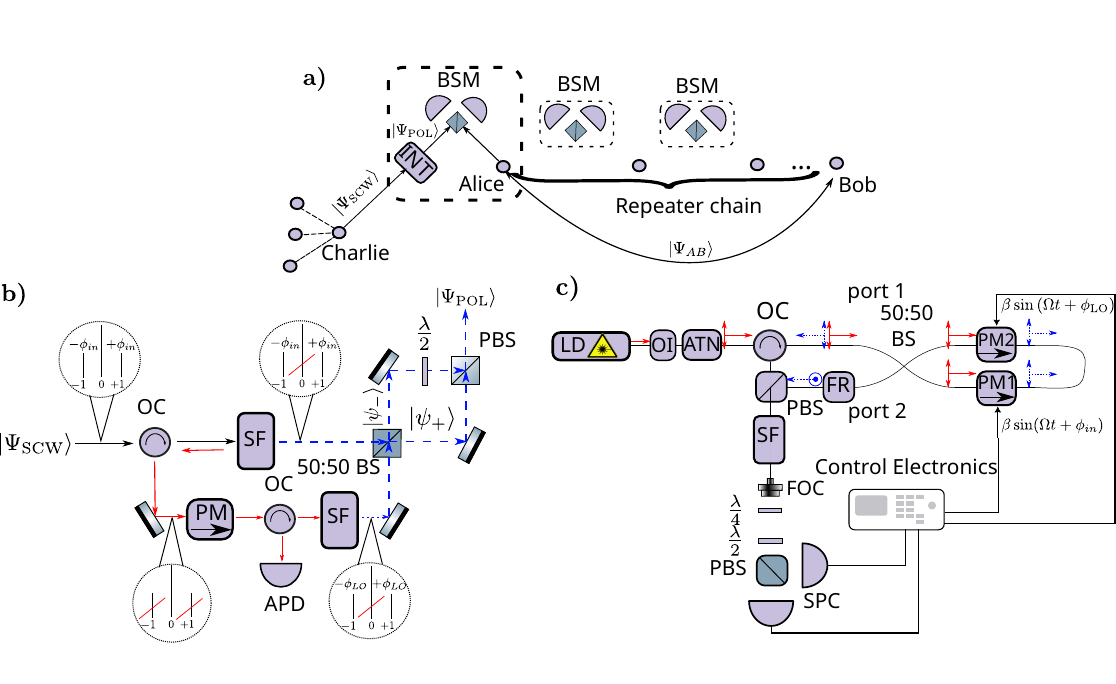}
\caption{\textbf{a)} The possible application of the interface (INT) for teleporting information from Charlie's SCW state $\ket{\Psi_{\text{SCW}}}$ to Bob with a use of Bell state measurement device (BSM) and entangled state between Alice and Bob $\ket{\Psi_{AB}}$ being generated via quantum repeater chain.  
\textbf{b)} The principle scheme of the interface that consist of phase modulator (PM), spectral filter (SF), controlling avalanche photodetector (APD), non-polarizing beam splitter 50:50 (BS),  polarization beam splitter (PBS), optical circulator (OC). The $\{-1,0,1\}$ indices in the circles indicate the presence of the sidebands  with their phases.  \textbf{c)}
The experimental setup with laser diode (LD), optical isolator (OI),  optical attenuator (ATN), single photon counting module (SPC), Faraday rotator (FR), fiber output coupler (FOC), quarter wave plate ($\lambda/4$), half wave plate ($\lambda/2$). \textcolor{red}{Red} solid and \textcolor{blue}{Blue} dashed arrows indicate the carrier and sidebands direction, respectively. Double arrow and circled dot represent horizontal and vertical polarizations, respectively. 
}
\label{fig:scheme}
\end{figure*}

Conceptually, the interface between SCW and dual rail encondings may be used as it is sketched in Fig.\ref{fig:scheme}\textbf{a}. 
Alice and Bob generate a non-local polarization entangled Bell state $\ket{\Psi_{AB}}$ with the help of a quantum repeater, while Alice and Charlie are connected with an optical link.  
Charlie, who is located in a trusted node of the SCW-QKD network, sends his SCW state $\ket{\Psi_{\text{SCW}}}$ to Alice, where the state is routed to a polarization Bell state measurement device via the interface. 
The interface maps the useful information from $\ket{\Psi_{\text{SCW}}}$ into a polarization qubit's degree of freedom $\ket{\Psi_{\text{POL}}}$. 
In turn, the polarization qubit is interfered with Alice's part of the shared state. The proper detection at the Bell state measurement device together with local polarization rotation at Bob's part teleports Charlie's information to Bob \cite{Bouwmeester1997}.   
However, a scheme for an appropriate interface has not been addressed until now.

In this work, we propose and demonstrate a proof of principle experiment on interfacing the SCW with dual rail encodings. First, we describe the scheme of the interface and its simple theoretical model. Next, we portray the experimental setup and present its results. Finally, we discuss the perspectives and further possible improvements of the interface.

\section{Principle scheme}
In SCW, the information is encoded by applying a sinusoidal phase modulation with modulation depth $\beta$, modulation frequency $\Omega$, and phase $\phi_{in}$ to a coherent state with complex amplitude $\alpha_0$ and carrier frequency $\omega_0$.
An optical state after a phase modulator is described by wavefunction \cite{Miroshnichenko:17}:
\fla{
\ket{\Psi_{\text{SCW}}} = \bigotimes^{S}_{m=-S} \ket{ \alpha_0 J_{m}(\beta)e^{i m \phi_{in}} }_m, \label{eq::SCW-State}
}
where multiplication is carried over 2S+1 sidebands with $m\in \mathbb{Z}$ being indices of sidebands with frequencies $ \omega_0 + m\Omega$, $J_{m}(\beta)$ is m-th order Bessel function of the first kind. 
The value of the phase $\phi_{in}$ encodes useful information.
For example, in typical SCW-QKD with BB-84-like protocol, Alice sends to Bob logical bits in two non-orthogonal bases with a single basis formed by pairs of states (\ref{eq::SCW-State}) with $\phi_{in}$ being shifted by $\pi$ with respect to each other, such as ${0,\pi}$ and ${\pi/2,3\pi/2}$.
In turn, Bob measured Alice's state in given basis by phase modulating it with the same $\beta$ and $\Omega$, but his own subcarrier phase and performing photon detection on sidebands ($m\neq0$) that effectively acts as a projection on a given state \cite{Miroshnichenko:17}.   
For interfacing the state (\ref{eq::SCW-State}) with some dual encoding, the phase $\phi_{in}$ has to be mapped into a superposition of two modes.

The principle scheme of the interface is depicted in Fig. \ref{fig:scheme}\textbf{b}. 
The input state $\ket{\Psi_{\text{SCW}}}$ with horizontal polarization enters spectral filter that reflects the carrier component ($m=0$) with power coefficient $r$ and transmits it  with power coefficient $1-r$. 
Meanwhile the sidebands ($m\neq0$) are transmitted through the filter with power coefficient $1-\varrho$ and reflected with power coefficient $\varrho$, where $\varrho$ is called as filter's sideband suppression factor.
The reflected carrier is routed by an optical circulator to a phase modulator, that modulates the carrier with the same modulation depth $\beta$, modulation frequency $\Omega$, but a fixed phase $\phi_{\text{LO}}$. 
After the phase modulator another spectral filter with the same $r$ and $\varrho$ separates the carrier frequency component from the generated sidebands.

The sidebands from the original state and the ones generated locally are overlapped on the 50:50 beam splitter.  
The wavefunction for two spatial modes at the output of the beam splitter labelled as '$+$' and '$-$' is
\fla{
\ket{\Psi} = \ket{\psi_{+}} \otimes \ket{\psi_{-}}, 
}
with 
\fla{
\ket{\psi_{\pm}} = \ket{\alpha_0\frac{J_0(\beta)((1-r) \pm J_0(\beta)r(1-r)) }{\sqrt{2} }}_0  \cdot \nonumber \\ 
\bigotimes_{m\neq0} \ket{\alpha_0  J_m(\beta) (1-\varrho)  \frac{  e^{im\phi_{in}} \pm  e^{im\phi_{\text{LO}}} r J_0(\beta) }{\sqrt{2}}}_m . \label{eq::dr}
}
For good enough filters ($r\approx 1$, $\varrho \approx 0$) and small modulation depth ($J_0(\beta)\approx 1$), the state in Eq. \eqref{eq::dr} represents the dual rail encoding with two spatial modes for each individual sideband with $m\neq 0$, e.g., the states with a phase difference of 0 or $\pi$ with respect to $\phi_{LO}$ will follow different channels of the beam splitter.
Both modes are combined on the polarization beam splitter, with one mode's polarization being rotated by $90^{o}$ with the half-waveplate and the mode's phase being shifted by $e^{i\phi}$ that is defined by the optical path difference.   
The resulted state for the given first order sidebands ($m=\pm 1$) up to a first order in photon number is:
\fla{
\ket{\Psi_{\pm 1}} \approx 
\ket{0_H 0_V}_{\pm1} + \sqrt{2}\alpha_0 J_{\pm1}(\beta) e^{\pm i(\phi_{in}+\phi_{\text{LO}}) /2 }\cdot   \nonumber \\
 \bigg(\cos \frac{\phi_{in}-\phi_{\text{LO}}}{2} \ket{1_H 0_V}_{\pm1} \pm  \sin\frac{\phi_{in}-\phi_{\text{LO}}}{2} \ket{0_H 1_V}_{\pm1} \bigg),
 \label{eq::pstate}
}
where we set the phase shift of V mode to be $e^{i\phi}=-i$. Once $\phi_{\text{LO}}$ is fixed to 0, the phase information $\phi_{in}$ is mapped on the probability amplitude of a photon being in a given polarization mode, as in dual rail encoding.

\begin{figure}
\centering
\includegraphics[width=0.6\linewidth]{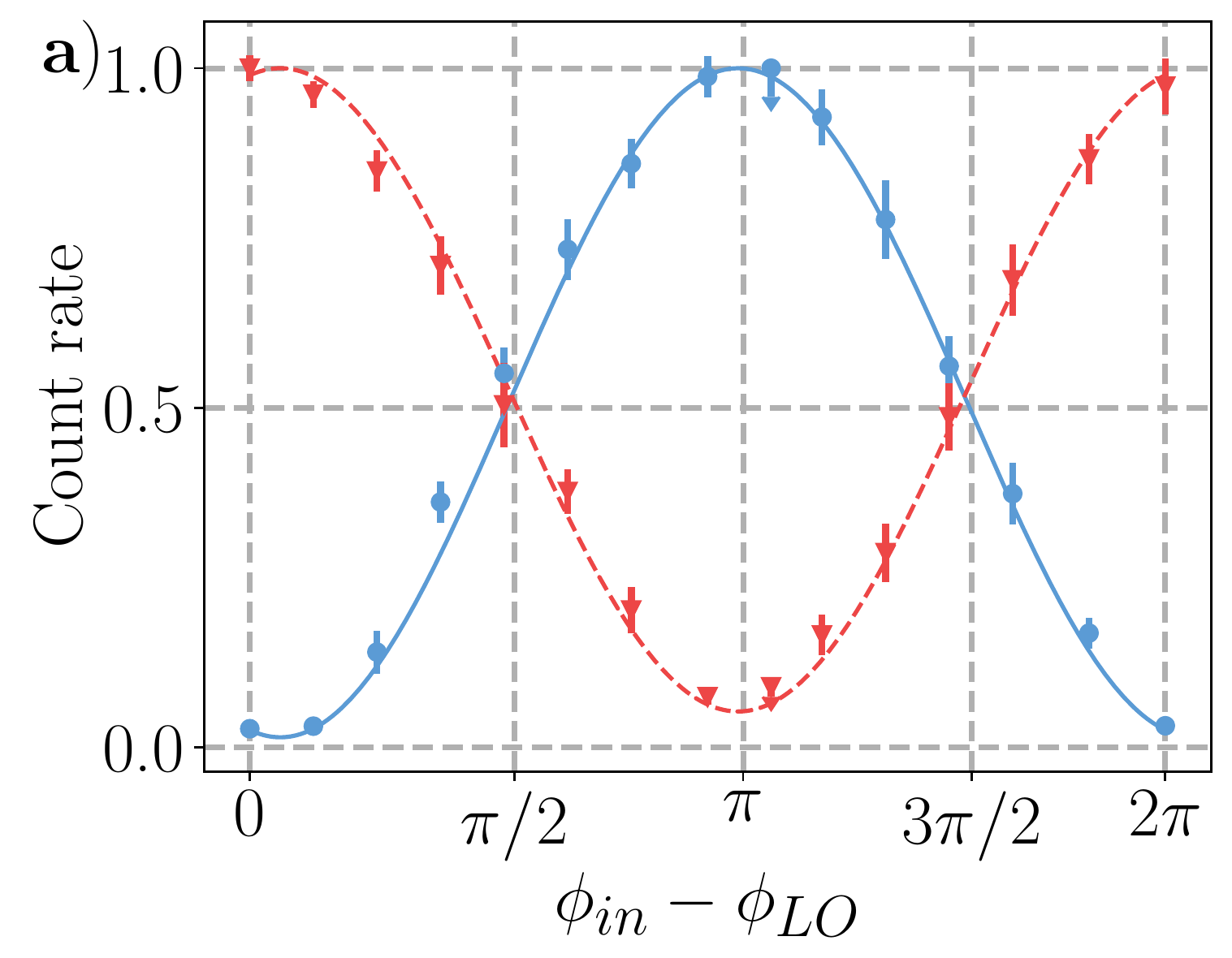} \\
\includegraphics[width=0.6\linewidth]{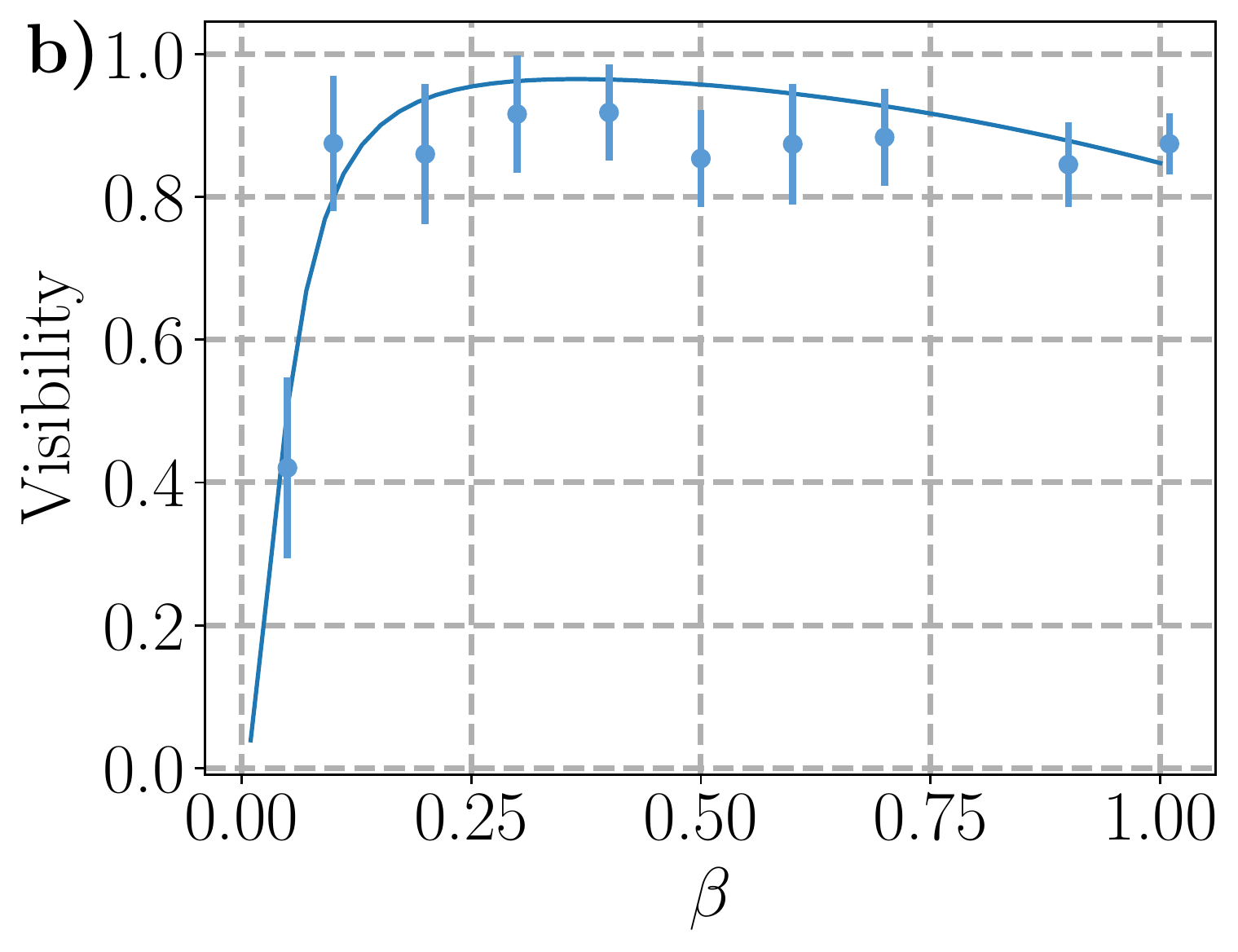}
\caption{ \textbf{a)} Dependence of the normalized count rate for $\ket{H}$ (\textcolor{red}{red} triangles) and $\ket{V}$ (\textcolor{blue}{blue} dots) states at different values of phase difference between two microwave generators. The data is obtained with modulation depth $\beta=0.15$. The dashed and solid lines represent theoretical fits.
\textbf{b)} Dependence of interference visibility on modulation depth. The solid curve is a theoretical fit. 
}
\label{fig:interference}
\end{figure}

\begin{figure}[ht]
\centering
\includegraphics[scale=.25]{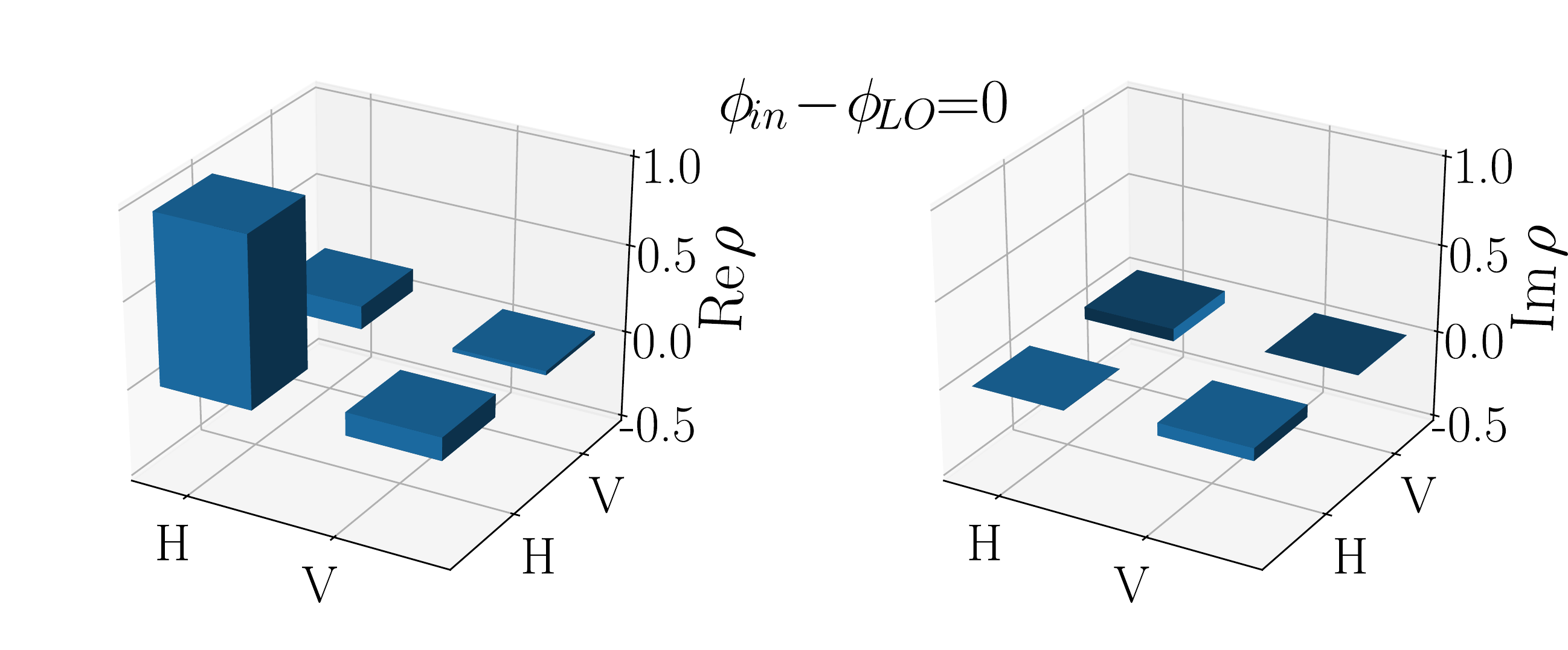} 
\includegraphics[scale=.25]{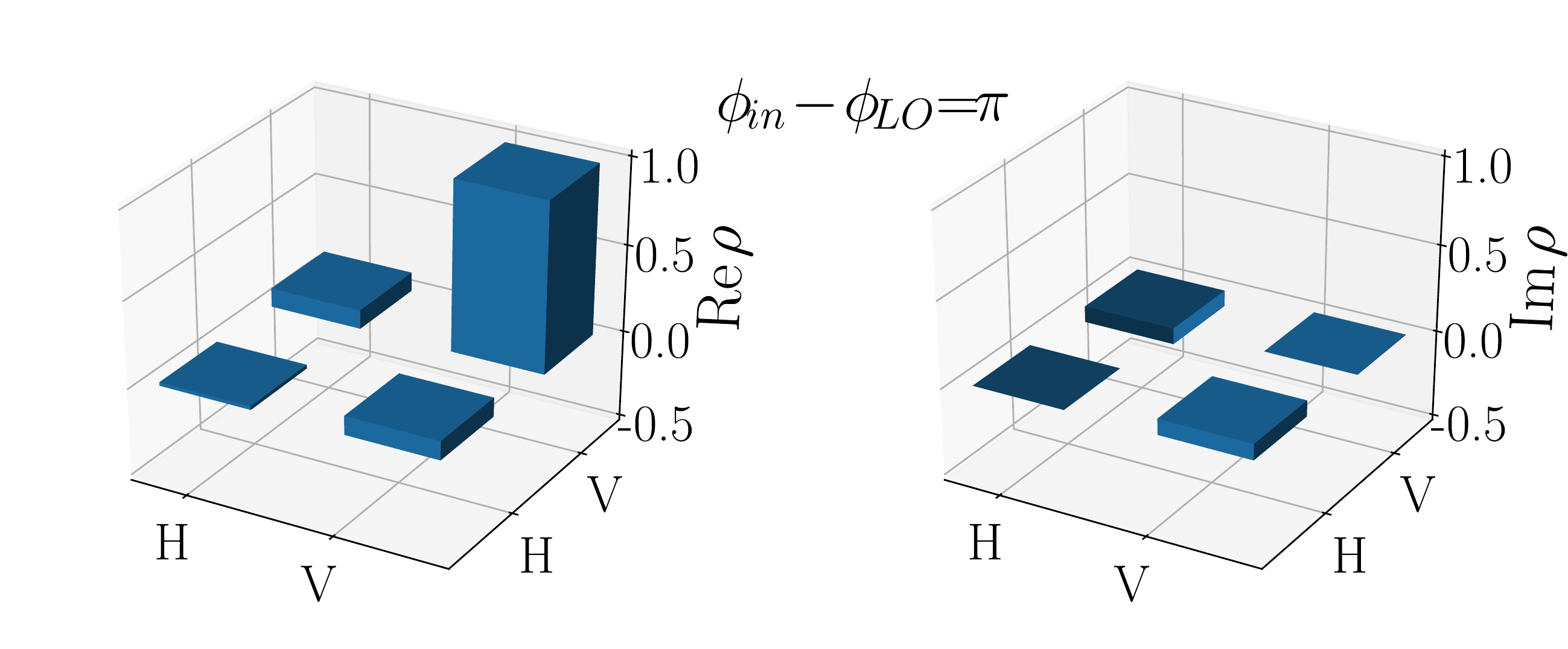} \\
\includegraphics[scale=.25]{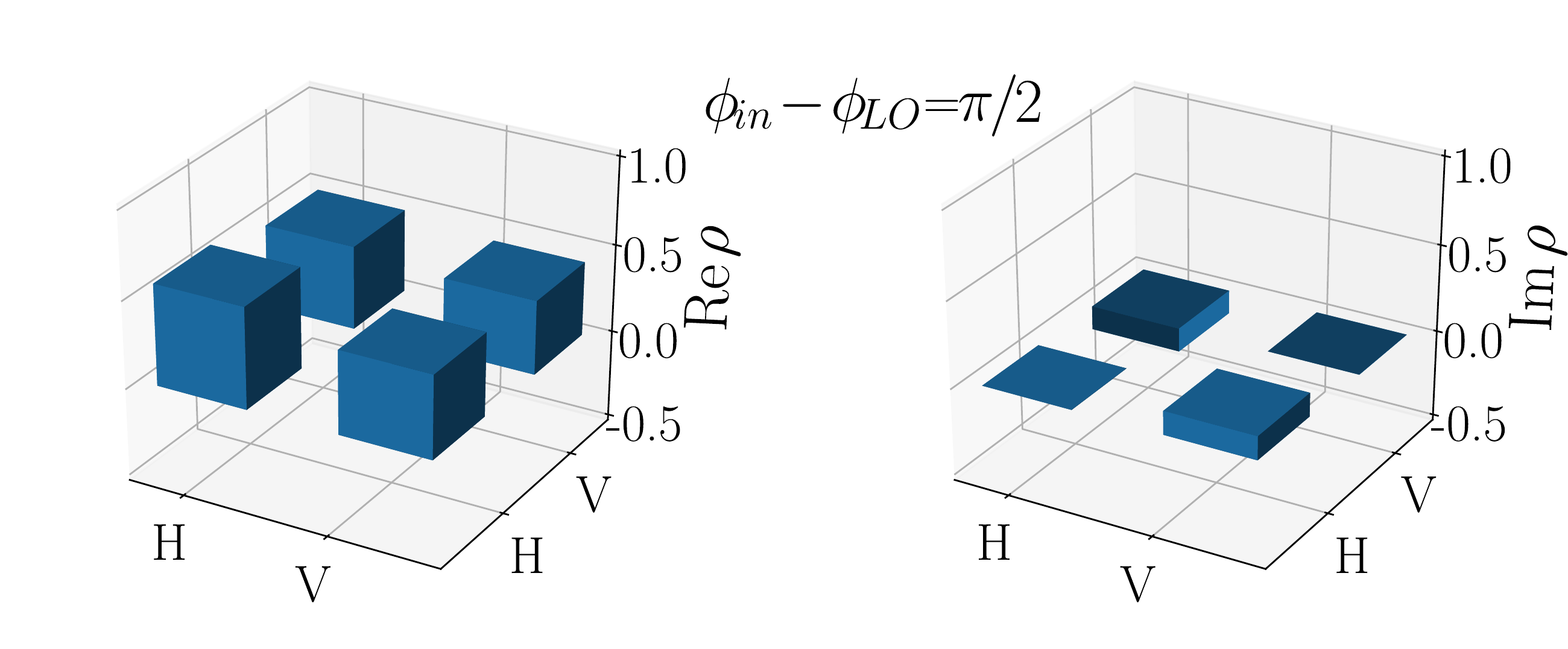}
\includegraphics[scale=.25]{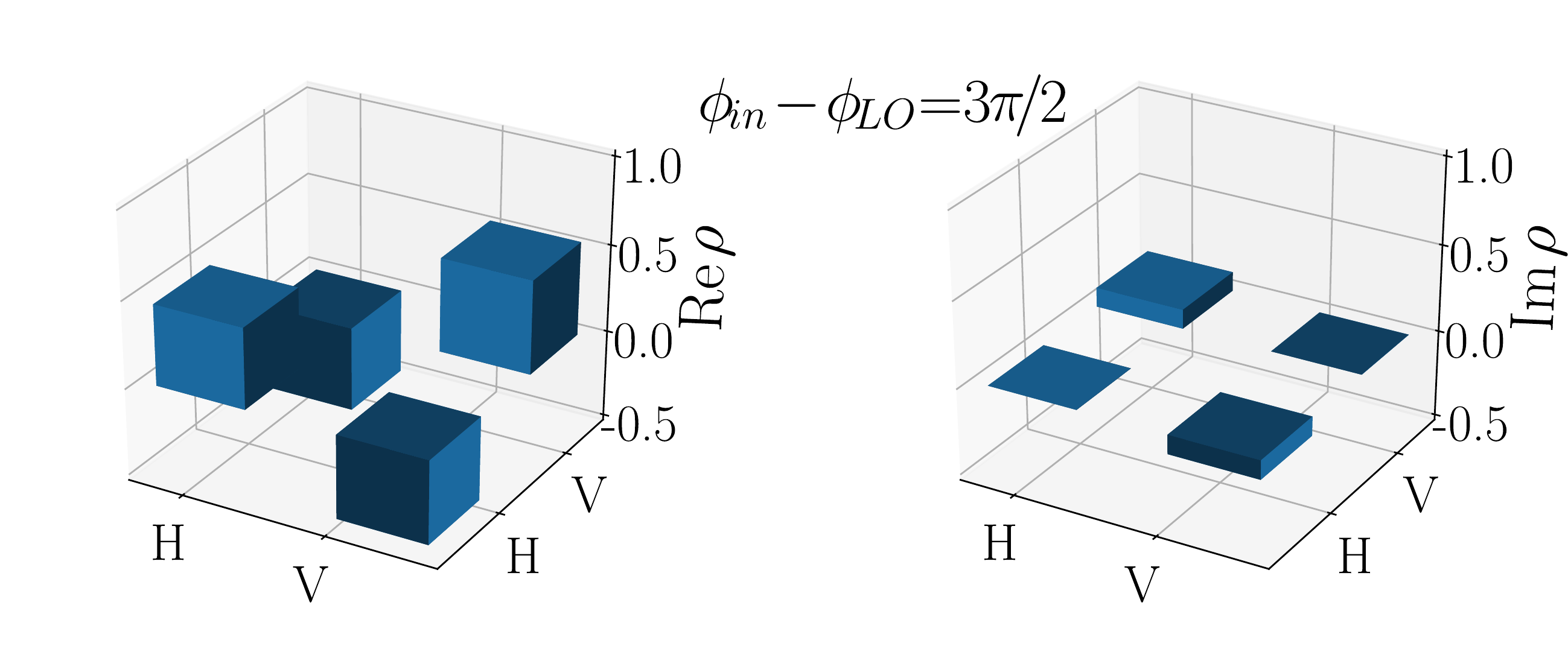} 
\caption{ Real and imaginary parts of the reconstructed density matrices for polarization states at different values of the phase difference in $\{\ket{H},\ket{V}\}$ basis.}
\label{fig:tomography}
\end{figure}

\section{Experiment}
The experimental scheme is depicted in Fig. \ref{fig:scheme}.
The fiber-bragg grating stabilised laser diode (TeraXion PureSpectrum LM) protected by an optical isolator together with an optical attenuator are used to generate continuous wave at 1550 nm.  
The laser is guided through an optical circulator into a 4-port fiber 50:50 beam splitter (BS).
Two SCW states are generated using the same carrier in a Sagnac-type interferometer, where clockwise and counter-clockwise travelling waves  are modulated by two separate $\text{LiNb}\text{O}_3$ phase modulators, PM1 an PM2, respectively.
The phase modulators are driven by two phase-locked microwave generators with an identical frequency of $\Omega=2\pi\cdot 4.8$ GHz but different phases, $\phi_{\text{in}}$ and $\phi_{\text{LO}}$, respectively. 
The implemented design helps to circumvent a problem with the need for two identical spectral filters, which is necessary for the principle scheme in Fig. \ref{fig:scheme}\textbf{b}. 

The generated optical sidebands are interfered on 50:50 BS and routed to ports 1 and 2 according to their phase difference.
The carrier component, meanwhile, is routed back to the incoming port 1 with isolation greater than 30 dB due to constructive interference between clockwise and counter-clockwise waves in the Sagnac loop. 
The polarization of the sidebands travelling from port 2 is rotated by 90 degrees with the fiber Faraday rotator.
The radiation from port 1 is guided by the optical circulator to a fiber polarization beam splitter (PBS), where this radiation is combined with sidebands from port 2 having orthogonal polarization.
The polarization insensitive fiber Bragg grating spectral filter (TeraXion OF) with r=0.99 and $\rho=10^{-4}$ separates the prepared polarization state of the sidebands from the carrier.
The light is sent to free space via a fiber output coupler for further analysis.
The polarization of the sidebands is characterized with a conventional polarization analyzer consisting of quarter and half waveplates together with PBS, where both channels are coupled into single photon counting modules (ID Quantique ID210) with quantum efficiency of $\epsilon=$10\% and dark count rate of $\gamma\sim$100 Hz. The detectors are gated at 100 MHz with a gate duration of $\delta t=3.3$ ns. For this time window, 1.0 pW of incident carrier power corresponds to a value of $\alpha_0 \sim 0.15$ that results in a maximal count rate of $\sim 10^{4}$ cps at single sideband.
Multi-channel frequency counter is used to monitor the count rates from both single photon counting modules.

For estimating the conversion rate, we define the efficiency as the ratio between the number of photons in sidebands with $m\neq 0$ before and after the interface.
The power efficiency of the conversion is determined by the highest optical losses out of the losses in the two channels of the interferometer. The identical losses are necessary for the indistinguishability of coherent states interfering on a 50:50 beam splitter.
The loss budget is composed of insertion losses in the phase modulator (-3.1 dB), optical circulator (-0.7 dB), non-polarizing beam splitter 50:50 (-0.5 dB), Faraday rotator (-0.6 dB) and  polarization beam splitter (-0.2 dB). The losses of -4.5 dB in both channel correspond to the efficiency of 35\%. 

\section{Results and Discussion}
First, we scan $\phi_{\text{in}}-\phi_{\text{LO}}$ between two SCW states by manually changing phase of a single microwave generator. The normalized count rate of horizontally and vertically polarized photons  at different value of the phase difference  are presented in Fig. \ref{fig:interference}a, where each point is accumulated during $T=10$ s.
The interference with visibility of $\sim95\%$ clearly indicates the transfer of the phase information from SCW state into polarization degree of freedom according to \eqref{eq::pstate}. 

We measure visibility at different values of modulation depth $\beta$ with all sidebands being detected simultaneously. 
The results are depicted in Fig \ref{fig:interference}b. 
At a small value ($\beta<0.1$) of modulation depth, the number of detected photons drops below the dark count rate that leads to low visibility.
At a large value of $\beta$ ($\beta>1$) the significant portion of energy is distributed at higher order sidebands ($m\ge2$) that interfere as $\sim \cos (\frac{m(\phi_{in}-\phi_{\text{LO}})}{2})$  and reduce overall visibility when all sidebands are detected simultaneously by a single detector.
At the same time, the visibility of $92 \pm 3\%$ is achievable within the range of $0.1 \le \beta\le0.7$.  
It is worth noting that the measured decrease in visibility at low $\beta$ is not fundamental but rather implementation specific and can be circumvented, if a detector with a lower dark count rate and a spectral filter with a larger extinction ratio $r/(1-r)$ are used.
In principle, the negative effect of the higher order ($\abs{m}>1$) sidebands can be mitigated by separating the first order sideband with an additional spectral filter. However, the additional filter would introduce extra losses and proportionally reduce the conversion efficiency.

We model the number of clicks with given polarization as:
\fla{
n_{H(V)} \approx \left(\gamma + \frac{\epsilon}{\delta t}\bra{\psi_{+(-)}} n \ket{\psi_{+(-)}}  \right) \cdot T,
\label{eq::clicks}
}
where $n  = \sum_{m \neq 0} \hat{a}^{\dagger}_m \hat{a}_m $ is number operator with $\hat{a}_m$ ($\hat{a}^{\dagger}_m$) being annihilation (creation) operator for m'th spectral mode, $\epsilon$ is the quantum efficiency of the detector, $\delta t$ is gate duration time, $T$ is sampling time, $\gamma$ is dark count rate. 
The experimentally acquired data of normalized clicks and visibility is fitted using  \eqref{eq::clicks}.

The polarization states on the opposite sidebands with m=1 and m=-1 have a $\pi$ phase shift with respect to their polarization states, which results in a mixed polarization for $\phi_{in}-\phi_{\text{LO}} \in \{\pi/2,3\pi/2\}$ if both sidebands are measured simultaneously.
To characterise the conversion, we perform the polarization tomography for only a single sideband ($m=+1$) with $\beta=0.15$ by adjusting the laser's frequency with respect to the spectral filter's reflection band such that only one sideband is transmitted. For each value of phase difference ($0,\pi,\pi/2,3\pi/2$) we measure polarization in different polarization basis. 
The acquired data is used to reconstruct the polarization density matrix by the maximum likelihood method \cite{Rehacek2001}. The reconstructed density matrices are presented in Fig. \ref{fig:tomography} with corresponding fidelities summarised in Table \ref{table:fidelities}.

The small imaginary part of non-diagonal elements of the density matrix we attribute to a jitter in optical path difference between paths for $\ket{H}$ (port 1 to PBS) and $\ket{V}$ (port 2 to PBS) photons. 
The jitter introduces extra complex phase between $\ket{H}$ and $\ket{V}$ that leads to unwanted ellipticity of the polarization. 
We passively stabilized the optical scheme by placing it in thermally and acoustically isolating foam and reducing the fiber connection length. 
However, some sort of active stabilization may still be required for achieving fidelity $\ge 99$\%. It must be noted, that typical teleporation fidelity is around $89\pm 2\%$ with current technologies \cite{PRXQuantum.1.020317}. Hence, the interface with fidelity $\ge 99\%$ should not limit the teleportation in practical applications.

\begin{table}[htp]
\centering
\begin{tabular}{|c|c|c|}
\hline
$\phi_{in}-\phi_{LO}$ & Polarization state  & Fidelity \\ \hline
$0$ & $\ket{H}$  &  $96\pm3$ \% \\ \hline
$\pi$ & $\ket{V}$ & $96\pm3$ \%  \\ \hline
$\pi/2$ & $(\ket{H}+\ket{V})/\sqrt{2}$  & $95\pm3$\% \\ \hline
$3\pi/2$ & $(\ket{H}-\ket{V})/\sqrt{2}$ & $95\pm3$\% \\  \hline 
\end{tabular}
\caption{The correspondence between the phase and polarization with the achieved fidelities.}
\label{table:fidelities}
\end{table}

The proposed interface may be implemented in integrated quantum optical fashion \cite{Sibson2017}. 
This approach would minimize the phase fluctuations, device size and hence increase its stability. Moreover, this technology would allow to build two identical spectral filters for effective conversion of the external SCW state as in the proposed scheme. 
The lithium niobate on insulator platform seems a promising candidate for implementing the proposed interface. The principle components of the interface such as compact phase modulators \cite{Wang2018}, high finesse microring resonator \cite{zhang2017monolithic}, polarizing and non-polarizing beam splitter \cite{Boes2018} were demonstrated.

In this case, this scheme can be used in SCW-QKD for simultaneous detection of two states in a single basis. 
If single photon detectors are placed at the output of each channel of a 50:50 beam splitter, the states with phases $\phi_{\text{LO}}$ and $\phi_{\text{LO}}+\pi$ are distinguishable by a click of the corresponding detector.
In the appendix, we compare the theoretical asymptotical key rate of the QKD with the interface used as two-orthogonal state discriminator and the conventional approach \cite{Miroshnichenko:18} with a single state detection. 
Simulation shows that use of the proposed scheme in point-to-point SCW encoded QKD may increase the key rate by about twice compared to the conventional approach.

It is worth mentioning that the potential eavesdropper may try to corrupt the security of the communication between Charlie and Alice by photon number splitting or unambiguous state discrimination attacks on multiphoton components \cite{HOROSHKO20191728}. For this reason, the carrier after the phase modulator should be monitored by a photodetector (APD in Fig. \ref{fig:scheme} \textbf{b}) for detecting the attacks with the strong reference method \cite{Koashi04}. This countermeasure imposes a condition of trustworthiness only on Alice's node in the quantum repeater but not the whole quantum repeater chain, which is a reasonable cost for interconnecting the heterogeneous network. Alternatively, one can implement additional phase states for detecting the attacks \cite{Gaidash:22}.
At the same time, the number resolved detectors in a Bell-state measuring device may mitigate the multiphoton issues in quantum teleportation that occur with a practical source \cite{Krovi2016}.

In summary, we presented the way for conversion of phase information from SCW into the dual rail encoding. 
The experimentally realized in-fiber conversion from SCW into polarization encoding shows fidelities above 92\%.
If the first-order sideband is additionally filtered from the higher-order sidebands, the main limitations for further increasing visibility are the extinction ratio of the spectral filter and interference stability.
The proposed interface opens one more way for connecting existing quantum networks with different physical layers and increasing QKD distance and rate.
The reverse conversion from dual rail into SCW is also of particular interest for long-range transmission. However, this topic may be a part of a separate study since it requires mapping from discrete to continuous variables, which usually demands non-classical resources and heralded preparation as in \cite{Ulanov2017,Sychev2018}.

\begin{acknowledgments}
Authors appreciate support within framework project \# 00075-02-2020-051/1 from 02.03.2020. Authors profound thank S.A. Moiseev for fruitful discussions. We are grateful to M.M. Minnegaliev and M.A. Smirnov for lending us optomechanics and fiber-optics.
\end{acknowledgments}

\appendix
\section{Key rate comparison}

\begin{figure*}
\centering
\includegraphics[width=0.9\linewidth]{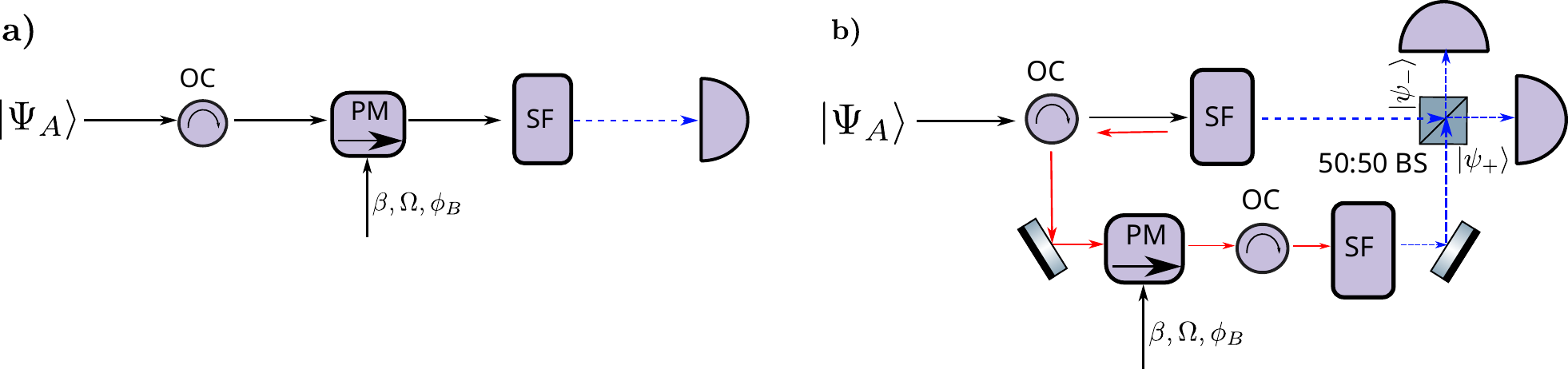} \quad 
\caption{ \textbf{a)} The traditional one-way SCW encoded QKD.  
\textbf{b)} The use of the proposed interface in one-way SCW encoded QKD for increasing key generation rate. The principle schemes consist from phase modulator (PM), spectral filter (SF), beam blocker (BB), non-polarizing beam splitter 50:50 (BS), optical circulator (OC). \textcolor{red}{Red} solid and \textcolor{blue}{Blue} dashed arrows indicate the carrier and sidebands direction, respectively.
} 
\label{fig:sketch}
\end{figure*}

\begin{figure*}
\centering
\includegraphics[width=0.35\linewidth]{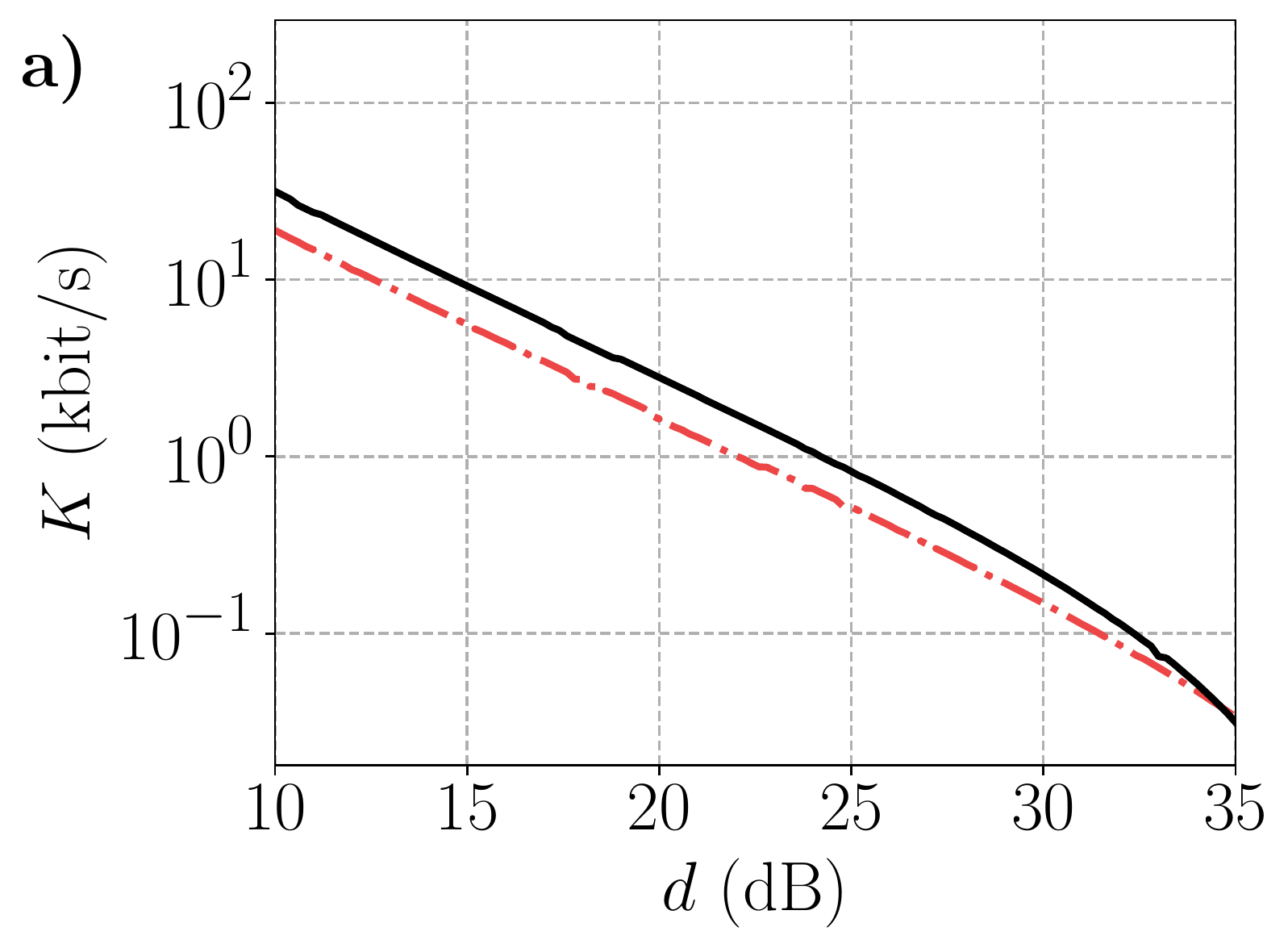} \quad \includegraphics[width=0.35\linewidth]{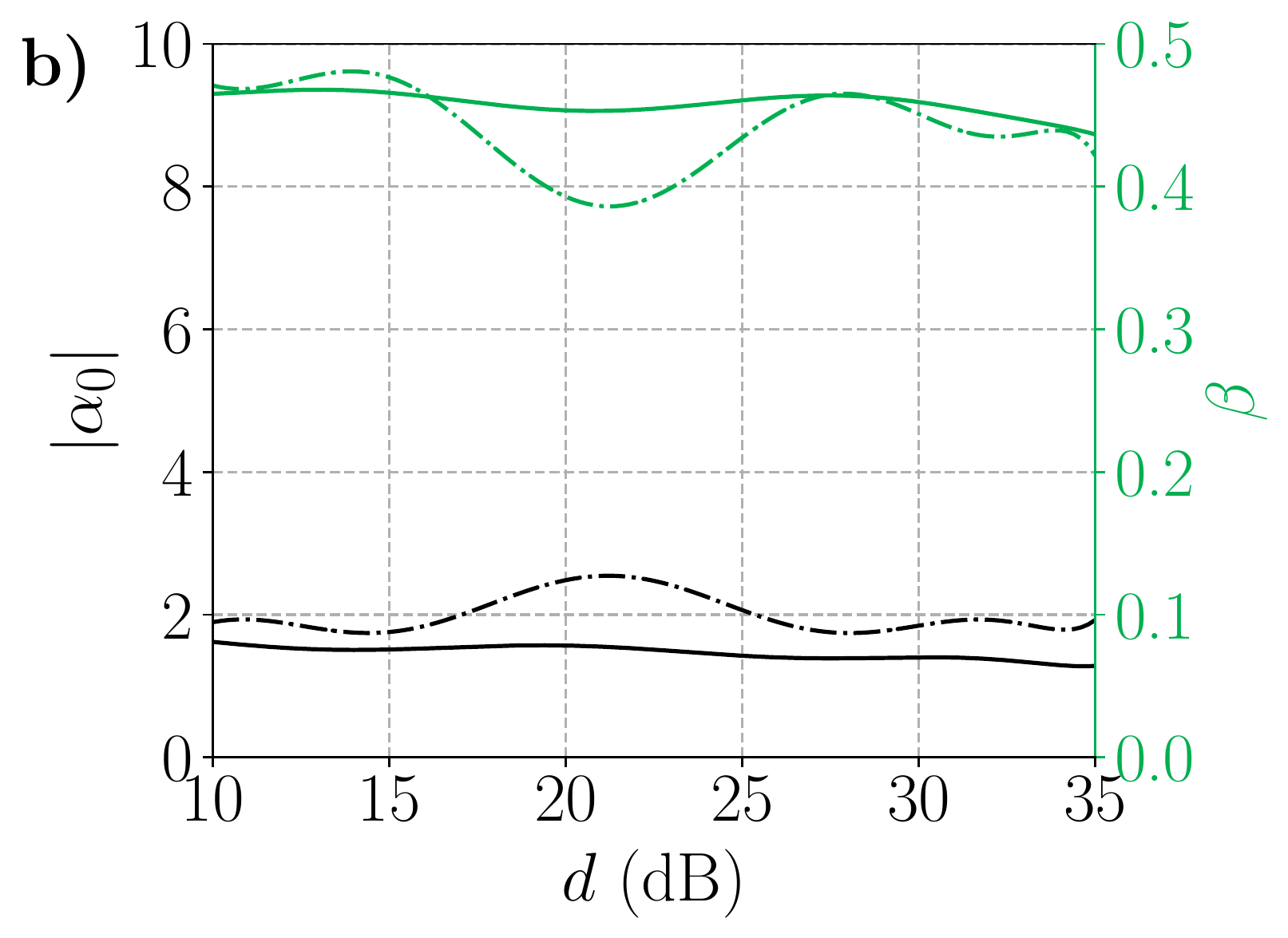} \\
\includegraphics[width=0.35\linewidth]{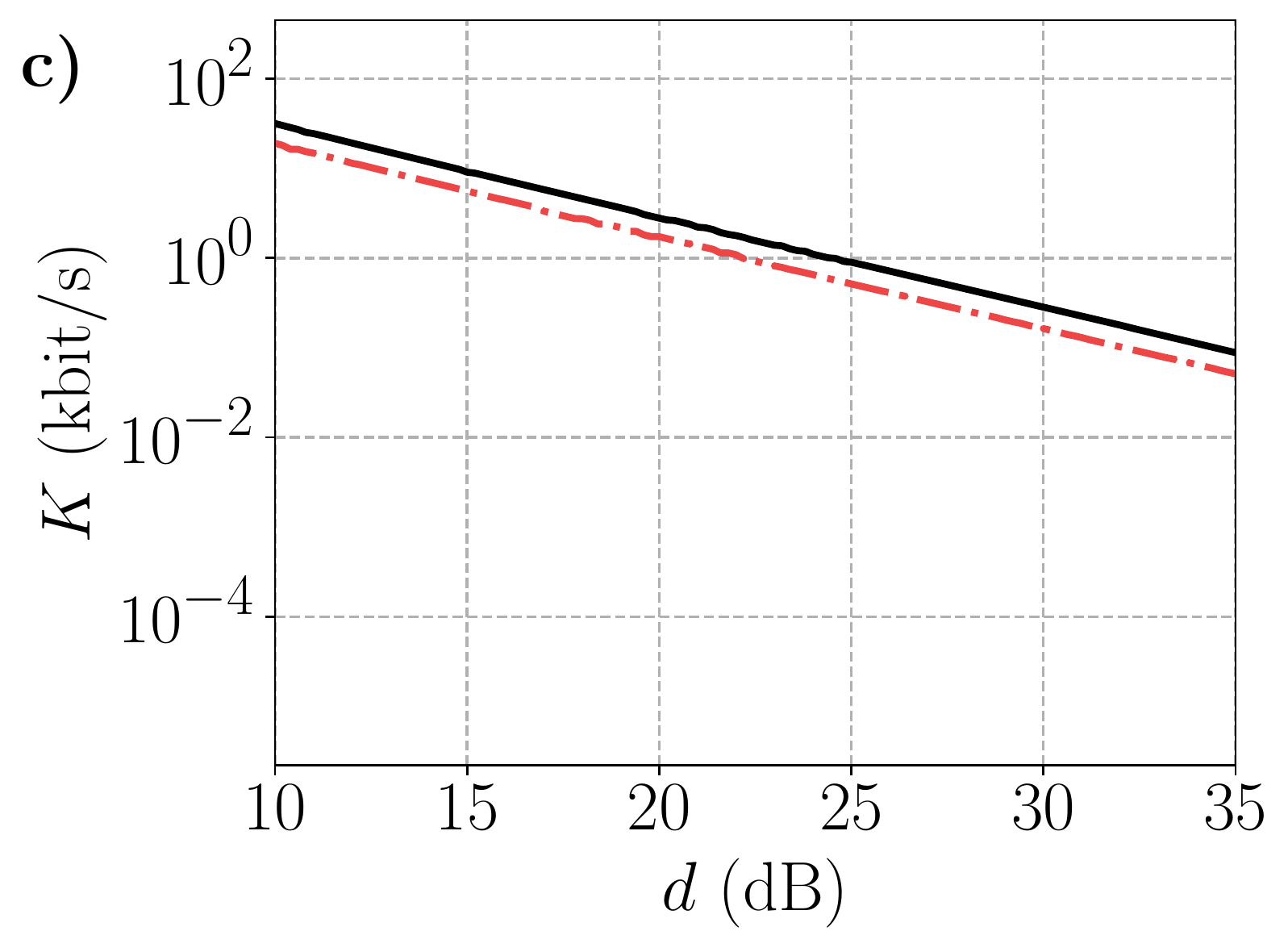} \quad \includegraphics[width=0.35\linewidth]{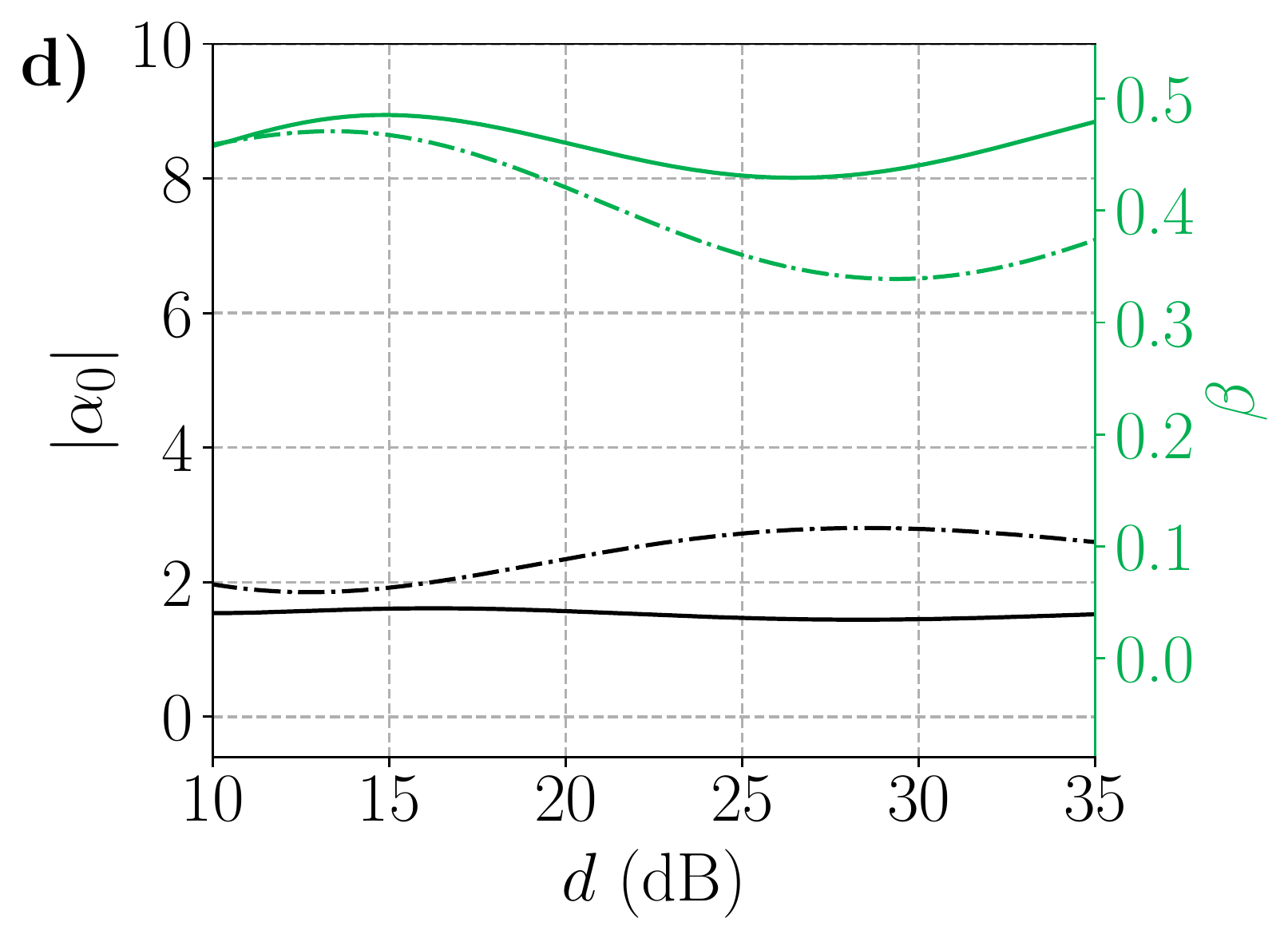}
\caption{ \textbf{a)} Simulated key generation rate $K$ of point-to-point SCW encoded QKD with the proposed interface used as state discrimination (\textbf{black} solid) and conventional detection system (\textcolor{red}{red} dashed).  
\textbf{b)} Optimal values of amplitude $\abs{\alpha_0}$ (\textbf{black} bottom) and modulation depth $\beta$ (\textcolor{green}{green} upper) for two-state discrimination (solid) and single discrimination (dashed). 
\textbf{c)} Simulated key generation rate $K$ with the same parameters and labeling as in a) but with $\gamma=1$ Hz. \textbf{d)} Optimal values of amplitude and modulation depth with labeling as in b) for $\gamma=1$ Hz.
}
\label{fig:KR}
\end{figure*}

Next we present the theoretical comparison of secret key generation rate with BB84 between traditional one-way SCW encoded QKD setup and SCW encoded QKD with the proposed interface being used as two logical states discriminator.

\subsection{Traditional scheme}
The scheme of the original one-way SCW encoded QKD is depicted in Fig.\ref{fig:sketch}\textbf{a}. Here Alice sends trains of pulses to Bob via an optical channel with losses $\eta$. 
Each of the pulses is in SCW state, with subcarrier phase $\phi_A$ being randomly set by Alice out of two sets that correspond to different bases. The Alice's state upon arrival to Bob is 
\fla{
\ket{\Psi_{A}} = \bigotimes^{S}_{m=-S} \ket{ \sqrt{\eta} \alpha_0 J_{m}(\beta)e^{i m \phi_{A}} }_m, 
}
where all parameters are described in section ``The principle scheme of the interface" of the main text.

At his side, Bob applies phase modulation to the received state with the same modulation depth $\beta$ but a randomly chosen value for the subcarrier phase $\phi_B$. 
The modulated state is transmitted through the spectral filter that separates the carrier ($m=0$) from the sidebands ($m\neq 0$). 
The sidebands are transmitted with power coefficient $1-\rho$, while the carrier is reflected with power coefficient $r$ and is transmitted with power coefficient $1-r$. After the filter, a non-discriminating single photon counter with quantum efficiency $\epsilon$ and a dark count rate $\gamma$ monitors the presence of a photon. 
At low intensity, the probability of the detector clicking is
\fla{
P_{\text{det}}(\phi_a,\phi_b) = \left( \epsilon \frac{n_{\text{ph}}(\phi_A,\phi_B) }{T} +\gamma  \right)\delta t, \label{eq:pdet}
}

where the $n_{\text{ph}} (\phi_A,\phi_B)$ is a number of photons after the spectral filter:
\begin{widetext}
\begin{flalign}
n_{\text{ph}} (\phi_A,\phi_B) = |\alpha_0|^2 \eta \cdot 
\begin{cases}
 (1-\varrho)(1-J^{2}_{0}(2\beta))+ (1-r) J^{2}_{0}(2\beta) \quad &\phi_A = \phi_B, \\
 (1-r) \quad &\phi_A = \phi_B+\pi, \\
\end{cases}
\label{eq:nph2}
\end{flalign}
\end{widetext}

where we assume negligibly small jitter between Alice's and Bob's modulation oscillators over a repetition period and the linear regime of the detector.

Asymptotic secret key generation rate against beam splitter collective attack  $K$ is lower limited by Devetak-Winter bound \cite{Devetak2005,Miroshnichenko:18}: 
\fla{
K = \nu P_B (1-  f(Q) H(Q) - \chi(A:E) ), \label{eq:keyrate}
}
where $\nu=1/T$ is a repetition rate and $P_B$ is probability of bit detection at Bob's side per pulse.  
Probability $P_B$ can be represented as a total probability of detecting a photon with correct and wrong bit decoding. Namely, it is a probability of the event that Alice encodes logical bit value x and Bob get the same bit value x or opposite bit value: $(x+1)\text{mod}~2$. 
For $x=0$ it is $P_B=(P_{\text{det}}(0,0)+P_{\text{det}}(0,\pi))/2$, where factor $1/2$ comes from the 2 bases used in BB84 protocol. 
The product $f(Q)H(Q)$ defines an amount of information revealed during error correction with $f(Q)$ being an efficiency of error correcting protocol and $H(Q)$ being a binary Shannon entropy:
\fla{
H(Q) = -Q \log_2 (Q) - (1-Q) \log_2(1-Q). \label{eq:entropy}
}
Here $Q$ is a quantum bit error probability (QBER), which is a ratio of detection with false decoding to the probability of detection $P_B$:
\fla{
Q = \frac{P_{\text{det}}(0,\pi)}{P_{\text{det}}(0,0)+P_{\text{det}}(0,\pi)}. \label{eq:qber}
}

The term $\chi(A:E)$ is Holevo bound for an amount of information that is accessible to an eavesdropper for a given collective attack.  
We assume Eve intercepts all photons lost in the optical channel between Alice and Bob and records them into her quantum memory unit. 
For sake of simplicity, we assume that Eve ``emulates'' losses by inserting a beam splitter with a power reflectivity coefficient $1-\eta$ at the beginning of the optical channel. Eve records the reflected photons into her memory, while she sends the transmitted photon to Bob via a lossless optical channel.
It was shown \cite{Miroshnichenko:18} that a collective beam splitter attack provides Eve with the following information:
\fla{
\chi(A:E) = H\left(\frac{1}{2}\left(1-e^{-|\alpha_0|^{2} (1-\eta) (1-J_0(2\beta)) } \right)\right).\label{eq:chi}
}
By combining \eqref{eq:chi} together with \eqref{eq:qber} and \eqref{eq:entropy} in \eqref{eq:keyrate} we simulate the secret key, assuming moderate efficiency of the error-correction protocol $f(Q)=1.25$ \cite{assche_2006}.

\subsection{Scheme with two state discriminator}

We consider the same one-way SCW-QKD  as in previous case but  with Bob being using the interface for measuring both state in a given bases during a single measurement. 
The sketch of the setup is presented in Fig. \ref{fig:sketch}\textbf{b}, where Alice's state is sent on the interface which output is subjected to single photon detection.   
As it is covered in the main text, the spectral filter separates the carrier frequency from the sidebands of the Alice's state. 
The carrier is phase modulated with subcarrier phase $\phi_B=\phi_{\text{LO}}$. 
The second spectral filter separates the generated sidebands from the carrier. 
The generated sidebands are overlapped at the 50:50 beamsplitter, whose output ports in modes ``$+$" and ``$-$" are monitored by a single photon counter with dark count rate $\gamma$.
The probabilities of the detectors in mode ``$+$" or ``$-$" getting a click are
\fla{
P_{\pm}(\phi_A,\phi_B) = \left( \epsilon \frac{n_{\pm}(\phi_A,\phi_B) }{T} +\gamma \right)\delta t, 
}
where 
\fla{
n_{\pm}(\phi_A,\phi_B) &= \sum_{m\neq 0} \abs{\sqrt{\eta} \alpha_0  J_m(\beta) (1-\varrho)  \frac{  e^{im\phi_A} \pm  e^{im\phi_{B}} r J_0(\beta) }{\sqrt{2}}}^2 \nonumber \\
&+ \abs{\sqrt{\eta} \alpha_0\frac{J_0(\beta)((1-r) \pm J_0(\beta)r(1-r)) }{\sqrt{2} }}^2. 
}
In the traditional scheme Bob discriminate particular state out of given basis by a particular choice of a subcarrier phase $\phi_B$ that corresponds to the target state.
In contrast, in the proposed scheme Bob can discriminate two state with subcarrier phase different by $\pi$ in a single measurement.  
Bob chooses the particular basis by setting the phase of local oscillator $\phi_{\text{LO}}$ of either 0 or $\pi/2$.  

QBER is calculated as a ratio between the probability of an erroneous bit decoding and the total probability of bit decoding from Alice to Bob
\fla{
Q = \frac{ P_E } {P_C + P_E },
\label{eq:dm-qber}
}
where probability of erroneous bit decoding is  
\fla{
P_E &= P_{+}(\phi_A, \phi_A+\pi )(1-P_{-}(\phi_A, \phi_A+\pi)) \nonumber \\ 
&+ P_{-}(\phi_A, \phi_A )(1-P_{+}(\phi_A, \phi_A ))
}
and probability of correct bit decoding is:
\fla{
P_C &= P_{+}(\phi_A, \phi_A )(1-P_{-}(\phi_A, \phi_A )) \nonumber \\ 
&+ P_{-}(\phi_A, \phi_A+\pi )(1-P_{+}(\phi_A, \phi_A+\pi )).
}
The probability of Bob detecting a bit erroneously or correctly is modified to $P_B = (P_E+P_C)/2$ accordingly. 
Since Alice's state has not changed, we may use the same Holevo bound for the eavesdropper as in the traditional case.  Hence, the  Eq.\eqref{eq:dm-qber} together with Eq.\eqref{eq:chi} are used in  Eq.~\eqref{eq:keyrate} to estimate the key generation rate \cite{codeSCWDL}. 

\begin{table}[htp]
\centering
\begin{tabular}{|c|c|}
\hline
Parameter & Description \\ \hline
$\epsilon=0.1$ & Detector quantum efficiency  \\ \hline
$\gamma=50$ Hz &  Detector's dark count rate \\ \hline
$r=0.99$ & Filter's carrier reflection \\ \hline
$\rho=-40$ dB & Filter's sideband suppression factor \\ \hline 
$\delta t=$3.3 ns & Gating time \\ \hline 
$T=$10 ns & Pulse repetition period \\ \hline 
\end{tabular}
\caption[Parameters used to find the noise spectrum]{Model parameters }
\label{table:parameters}
\end{table}

The key generation rates for the traditional and the proposed scheme are presented in Fig. \ref{fig:KR}\textbf{a} with the parameters being summarized in Table \ref{table:parameters}. For each value of $\eta$ the control parameters such as $\alpha_0$ and $\beta$ are numerically optimized. The optimized values are depicted in Fig. \ref{fig:KR}\textbf{b}. 
The proposed scheme demonstrates a key rate about twice as large as the conventional scheme at the given set of parameters. However, the higher-order sidebands ($\abs{m}\gg 1$) in the proposed scheme introduce an error proportional to $\alpha_0 \sum_{\abs{m}\ge 2} \abs{J_m(\beta)}^2$ that is absent in the conventional scheme.
It becomes especially clear when using a detector with lower dark count rates. 
The key rate for a model with the same parameters but $\gamma = 1$ Hz is depicted in Fig. \ref{fig:KR}\textbf{c}.  
At large values of losses (>50 dB), the conventional protocol apparently takes over due to the mentioned contribution of the higher-order sidebands. To reduce this effect, the optimal $\beta$ for the proposed scheme is decreased at larger losses, as shown in Fig. \ref{fig:KR}\textbf{d}. 
One way to circumvent this issue is to detect only one sideband, which may also help mitigate the chromatic dispersion \cite{Kiselev:20}.
It is worth noting, that we compare two QKD against the collective beam-splitter attack for a purpose of differential comparison only and use the same parameters for both case. 
For a complete security analysis more advanced attacks \cite{Sajeed2021} should be incorporated.

\bibliography{lib}

\begin{thebibliography}{41}%
\makeatletter
\providecommand \@ifxundefined [1]{%
 \@ifx{#1\undefined}
}%
\providecommand \@ifnum [1]{%
 \ifnum #1\expandafter \@firstoftwo
 \else \expandafter \@secondoftwo
 \fi
}%
\providecommand \@ifx [1]{%
 \ifx #1\expandafter \@firstoftwo
 \else \expandafter \@secondoftwo
 \fi
}%
\providecommand \natexlab [1]{#1}%
\providecommand \enquote  [1]{``#1''}%
\providecommand \bibnamefont  [1]{#1}%
\providecommand \bibfnamefont [1]{#1}%
\providecommand \citenamefont [1]{#1}%
\providecommand \href@noop [0]{\@secondoftwo}%
\providecommand \href [0]{\begingroup \@sanitize@url \@href}%
\providecommand \@href[1]{\@@startlink{#1}\@@href}%
\providecommand \@@href[1]{\endgroup#1\@@endlink}%
\providecommand \@sanitize@url [0]{\catcode `\\12\catcode `\$12\catcode
  `\&12\catcode `\#12\catcode `\^12\catcode `\_12\catcode `\%12\relax}%
\providecommand \@@startlink[1]{}%
\providecommand \@@endlink[0]{}%
\providecommand \url  [0]{\begingroup\@sanitize@url \@url }%
\providecommand \@url [1]{\endgroup\@href {#1}{\urlprefix }}%
\providecommand \urlprefix  [0]{URL }%
\providecommand \Eprint [0]{\href }%
\providecommand \doibase [0]{http://dx.doi.org/}%
\providecommand \selectlanguage [0]{\@gobble}%
\providecommand \bibinfo  [0]{\@secondoftwo}%
\providecommand \bibfield  [0]{\@secondoftwo}%
\providecommand \translation [1]{[#1]}%
\providecommand \BibitemOpen [0]{}%
\providecommand \bibitemStop [0]{}%
\providecommand \bibitemNoStop [0]{.\EOS\space}%
\providecommand \EOS [0]{\spacefactor3000\relax}%
\providecommand \BibitemShut  [1]{\csname bibitem#1\endcsname}%
\let\auto@bib@innerbib\@empty
\bibitem [{\citenamefont {Pirandola}\ \emph {et~al.}(2017)\citenamefont
  {Pirandola}, \citenamefont {Laurenza}, \citenamefont {Ottaviani},\ and\
  \citenamefont {Banchi}}]{Pirandola2017}%
  \BibitemOpen
  \bibfield  {author} {\bibinfo {author} {\bibfnamefont {S.}~\bibnamefont
  {Pirandola}}, \bibinfo {author} {\bibfnamefont {R.}~\bibnamefont {Laurenza}},
  \bibinfo {author} {\bibfnamefont {C.}~\bibnamefont {Ottaviani}}, \ and\
  \bibinfo {author} {\bibfnamefont {L.}~\bibnamefont {Banchi}},\ }\href
  {\doibase 10.1038/ncomms15043} {\bibfield  {journal} {\bibinfo  {journal}
  {Nature Communications}\ }\textbf {\bibinfo {volume} {8}},\ \bibinfo {pages}
  {15043} (\bibinfo {year} {2017})}\BibitemShut {NoStop}%
\bibitem [{\citenamefont {Elliott}(2002)}]{Elliott_2002}%
  \BibitemOpen
  \bibfield  {author} {\bibinfo {author} {\bibfnamefont {C.}~\bibnamefont
  {Elliott}},\ }\href {\doibase 10.1088/1367-2630/4/1/346} {\bibfield
  {journal} {\bibinfo  {journal} {New Journal of Physics}\ }\textbf {\bibinfo
  {volume} {4}},\ \bibinfo {pages} {46} (\bibinfo {year} {2002})}\BibitemShut
  {NoStop}%
\bibitem [{\citenamefont {Sasaki}\ \emph {et~al.}(2011)\citenamefont {Sasaki},
  \citenamefont {Fujiwara}, \citenamefont {Ishizuka}, \citenamefont {Klaus},
  \citenamefont {Wakui}, \citenamefont {Takeoka}, \citenamefont {Miki},
  \citenamefont {Yamashita}, \citenamefont {Wang}, \citenamefont {Tanaka},
  \citenamefont {Yoshino}, \citenamefont {Nambu}, \citenamefont {Takahashi},
  \citenamefont {Tajima}, \citenamefont {Tomita}, \citenamefont {Domeki},
  \citenamefont {Hasegawa}, \citenamefont {Sakai}, \citenamefont {Kobayashi},
  \citenamefont {Asai}, \citenamefont {Shimizu}, \citenamefont {Tokura},
  \citenamefont {Tsurumaru}, \citenamefont {Matsui}, \citenamefont {Honjo},
  \citenamefont {Tamaki}, \citenamefont {Takesue}, \citenamefont {Tokura},
  \citenamefont {Dynes}, \citenamefont {Dixon}, \citenamefont {Sharpe},
  \citenamefont {Yuan}, \citenamefont {Shields}, \citenamefont {Uchikoga},
  \citenamefont {Legr\'{e}}, \citenamefont {Robyr}, \citenamefont {Trinkler},
  \citenamefont {Monat}, \citenamefont {Page}, \citenamefont {Ribordy},
  \citenamefont {Poppe}, \citenamefont {Allacher}, \citenamefont {Maurhart},
  \citenamefont {L\"{a}nger}, \citenamefont {Peev},\ and\ \citenamefont
  {Zeilinger}}]{Sasaki:11}%
  \BibitemOpen
  \bibfield  {author} {\bibinfo {author} {\bibfnamefont {M.}~\bibnamefont
  {Sasaki}}, \bibinfo {author} {\bibfnamefont {M.}~\bibnamefont {Fujiwara}},
  \bibinfo {author} {\bibfnamefont {H.}~\bibnamefont {Ishizuka}}, \bibinfo
  {author} {\bibfnamefont {W.}~\bibnamefont {Klaus}}, \bibinfo {author}
  {\bibfnamefont {K.}~\bibnamefont {Wakui}}, \bibinfo {author} {\bibfnamefont
  {M.}~\bibnamefont {Takeoka}}, \bibinfo {author} {\bibfnamefont
  {S.}~\bibnamefont {Miki}}, \bibinfo {author} {\bibfnamefont {T.}~\bibnamefont
  {Yamashita}}, \bibinfo {author} {\bibfnamefont {Z.}~\bibnamefont {Wang}},
  \bibinfo {author} {\bibfnamefont {A.}~\bibnamefont {Tanaka}}, \bibinfo
  {author} {\bibfnamefont {K.}~\bibnamefont {Yoshino}}, \bibinfo {author}
  {\bibfnamefont {Y.}~\bibnamefont {Nambu}}, \bibinfo {author} {\bibfnamefont
  {S.}~\bibnamefont {Takahashi}}, \bibinfo {author} {\bibfnamefont
  {A.}~\bibnamefont {Tajima}}, \bibinfo {author} {\bibfnamefont
  {A.}~\bibnamefont {Tomita}}, \bibinfo {author} {\bibfnamefont
  {T.}~\bibnamefont {Domeki}}, \bibinfo {author} {\bibfnamefont
  {T.}~\bibnamefont {Hasegawa}}, \bibinfo {author} {\bibfnamefont
  {Y.}~\bibnamefont {Sakai}}, \bibinfo {author} {\bibfnamefont
  {H.}~\bibnamefont {Kobayashi}}, \bibinfo {author} {\bibfnamefont
  {T.}~\bibnamefont {Asai}}, \bibinfo {author} {\bibfnamefont {K.}~\bibnamefont
  {Shimizu}}, \bibinfo {author} {\bibfnamefont {T.}~\bibnamefont {Tokura}},
  \bibinfo {author} {\bibfnamefont {T.}~\bibnamefont {Tsurumaru}}, \bibinfo
  {author} {\bibfnamefont {M.}~\bibnamefont {Matsui}}, \bibinfo {author}
  {\bibfnamefont {T.}~\bibnamefont {Honjo}}, \bibinfo {author} {\bibfnamefont
  {K.}~\bibnamefont {Tamaki}}, \bibinfo {author} {\bibfnamefont
  {H.}~\bibnamefont {Takesue}}, \bibinfo {author} {\bibfnamefont
  {Y.}~\bibnamefont {Tokura}}, \bibinfo {author} {\bibfnamefont {J.~F.}\
  \bibnamefont {Dynes}}, \bibinfo {author} {\bibfnamefont {A.~R.}\ \bibnamefont
  {Dixon}}, \bibinfo {author} {\bibfnamefont {A.~W.}\ \bibnamefont {Sharpe}},
  \bibinfo {author} {\bibfnamefont {Z.~L.}\ \bibnamefont {Yuan}}, \bibinfo
  {author} {\bibfnamefont {A.~J.}\ \bibnamefont {Shields}}, \bibinfo {author}
  {\bibfnamefont {S.}~\bibnamefont {Uchikoga}}, \bibinfo {author}
  {\bibfnamefont {M.}~\bibnamefont {Legr\'{e}}}, \bibinfo {author}
  {\bibfnamefont {S.}~\bibnamefont {Robyr}}, \bibinfo {author} {\bibfnamefont
  {P.}~\bibnamefont {Trinkler}}, \bibinfo {author} {\bibfnamefont
  {L.}~\bibnamefont {Monat}}, \bibinfo {author} {\bibfnamefont {J.-B.}\
  \bibnamefont {Page}}, \bibinfo {author} {\bibfnamefont {G.}~\bibnamefont
  {Ribordy}}, \bibinfo {author} {\bibfnamefont {A.}~\bibnamefont {Poppe}},
  \bibinfo {author} {\bibfnamefont {A.}~\bibnamefont {Allacher}}, \bibinfo
  {author} {\bibfnamefont {O.}~\bibnamefont {Maurhart}}, \bibinfo {author}
  {\bibfnamefont {T.}~\bibnamefont {L\"{a}nger}}, \bibinfo {author}
  {\bibfnamefont {M.}~\bibnamefont {Peev}}, \ and\ \bibinfo {author}
  {\bibfnamefont {A.}~\bibnamefont {Zeilinger}},\ }\href {\doibase
  10.1364/OE.19.010387} {\bibfield  {journal} {\bibinfo  {journal} {Opt.
  Express}\ }\textbf {\bibinfo {volume} {19}},\ \bibinfo {pages} {10387}
  (\bibinfo {year} {2011})}\BibitemShut {NoStop}%
\bibitem [{\citenamefont {Dynes}\ \emph {et~al.}(2019)\citenamefont {Dynes},
  \citenamefont {Wonfor}, \citenamefont {Tam}, \citenamefont {Sharpe},
  \citenamefont {Takahashi}, \citenamefont {Lucamarini}, \citenamefont {Plews},
  \citenamefont {Yuan}, \citenamefont {Dixon}, \citenamefont {Cho},
  \citenamefont {Tanizawa}, \citenamefont {Elbers}, \citenamefont
  {Grei{\ss}er}, \citenamefont {White}, \citenamefont {Penty},\ and\
  \citenamefont {Shields}}]{Dynes2019}%
  \BibitemOpen
  \bibfield  {author} {\bibinfo {author} {\bibfnamefont {J.~F.}\ \bibnamefont
  {Dynes}}, \bibinfo {author} {\bibfnamefont {A.}~\bibnamefont {Wonfor}},
  \bibinfo {author} {\bibfnamefont {W.~W.-S.}\ \bibnamefont {Tam}}, \bibinfo
  {author} {\bibfnamefont {A.~W.}\ \bibnamefont {Sharpe}}, \bibinfo {author}
  {\bibfnamefont {R.}~\bibnamefont {Takahashi}}, \bibinfo {author}
  {\bibfnamefont {M.}~\bibnamefont {Lucamarini}}, \bibinfo {author}
  {\bibfnamefont {A.}~\bibnamefont {Plews}}, \bibinfo {author} {\bibfnamefont
  {Z.~L.}\ \bibnamefont {Yuan}}, \bibinfo {author} {\bibfnamefont {A.~R.}\
  \bibnamefont {Dixon}}, \bibinfo {author} {\bibfnamefont {J.}~\bibnamefont
  {Cho}}, \bibinfo {author} {\bibfnamefont {Y.}~\bibnamefont {Tanizawa}},
  \bibinfo {author} {\bibfnamefont {J.-P.}\ \bibnamefont {Elbers}}, \bibinfo
  {author} {\bibfnamefont {H.}~\bibnamefont {Grei{\ss}er}}, \bibinfo {author}
  {\bibfnamefont {I.~H.}\ \bibnamefont {White}}, \bibinfo {author}
  {\bibfnamefont {R.~V.}\ \bibnamefont {Penty}}, \ and\ \bibinfo {author}
  {\bibfnamefont {A.~J.}\ \bibnamefont {Shields}},\ }\href {\doibase
  10.1038/s41534-019-0221-4} {\bibfield  {journal} {\bibinfo  {journal} {npj
  Quantum Information}\ }\textbf {\bibinfo {volume} {5}},\ \bibinfo {pages}
  {101} (\bibinfo {year} {2019})}\BibitemShut {NoStop}%
\bibitem [{\citenamefont {Chen}\ \emph {et~al.}(2021)\citenamefont {Chen},
  \citenamefont {Zhang}, \citenamefont {Chen}, \citenamefont {Cai},
  \citenamefont {Liao}, \citenamefont {Zhang}, \citenamefont {Chen},
  \citenamefont {Yin}, \citenamefont {Ren}, \citenamefont {Chen}, \citenamefont
  {Han}, \citenamefont {Yu}, \citenamefont {Liang}, \citenamefont {Zhou},
  \citenamefont {Yuan}, \citenamefont {Zhao}, \citenamefont {Wang},
  \citenamefont {Jiang}, \citenamefont {Zhang}, \citenamefont {Liu},
  \citenamefont {Li}, \citenamefont {Shen}, \citenamefont {Cao}, \citenamefont
  {Lu}, \citenamefont {Shu}, \citenamefont {Wang}, \citenamefont {Li},
  \citenamefont {Liu}, \citenamefont {Xu}, \citenamefont {Wang}, \citenamefont
  {Peng},\ and\ \citenamefont {Pan}}]{Chen2021}%
  \BibitemOpen
  \bibfield  {author} {\bibinfo {author} {\bibfnamefont {Y.-A.}\ \bibnamefont
  {Chen}}, \bibinfo {author} {\bibfnamefont {Q.}~\bibnamefont {Zhang}},
  \bibinfo {author} {\bibfnamefont {T.-Y.}\ \bibnamefont {Chen}}, \bibinfo
  {author} {\bibfnamefont {W.-Q.}\ \bibnamefont {Cai}}, \bibinfo {author}
  {\bibfnamefont {S.-K.}\ \bibnamefont {Liao}}, \bibinfo {author}
  {\bibfnamefont {J.}~\bibnamefont {Zhang}}, \bibinfo {author} {\bibfnamefont
  {K.}~\bibnamefont {Chen}}, \bibinfo {author} {\bibfnamefont {J.}~\bibnamefont
  {Yin}}, \bibinfo {author} {\bibfnamefont {J.-G.}\ \bibnamefont {Ren}},
  \bibinfo {author} {\bibfnamefont {Z.}~\bibnamefont {Chen}}, \bibinfo {author}
  {\bibfnamefont {S.-L.}\ \bibnamefont {Han}}, \bibinfo {author} {\bibfnamefont
  {Q.}~\bibnamefont {Yu}}, \bibinfo {author} {\bibfnamefont {K.}~\bibnamefont
  {Liang}}, \bibinfo {author} {\bibfnamefont {F.}~\bibnamefont {Zhou}},
  \bibinfo {author} {\bibfnamefont {X.}~\bibnamefont {Yuan}}, \bibinfo {author}
  {\bibfnamefont {M.-S.}\ \bibnamefont {Zhao}}, \bibinfo {author}
  {\bibfnamefont {T.-Y.}\ \bibnamefont {Wang}}, \bibinfo {author}
  {\bibfnamefont {X.}~\bibnamefont {Jiang}}, \bibinfo {author} {\bibfnamefont
  {L.}~\bibnamefont {Zhang}}, \bibinfo {author} {\bibfnamefont {W.-Y.}\
  \bibnamefont {Liu}}, \bibinfo {author} {\bibfnamefont {Y.}~\bibnamefont
  {Li}}, \bibinfo {author} {\bibfnamefont {Q.}~\bibnamefont {Shen}}, \bibinfo
  {author} {\bibfnamefont {Y.}~\bibnamefont {Cao}}, \bibinfo {author}
  {\bibfnamefont {C.-Y.}\ \bibnamefont {Lu}}, \bibinfo {author} {\bibfnamefont
  {R.}~\bibnamefont {Shu}}, \bibinfo {author} {\bibfnamefont {J.-Y.}\
  \bibnamefont {Wang}}, \bibinfo {author} {\bibfnamefont {L.}~\bibnamefont
  {Li}}, \bibinfo {author} {\bibfnamefont {N.-L.}\ \bibnamefont {Liu}},
  \bibinfo {author} {\bibfnamefont {F.}~\bibnamefont {Xu}}, \bibinfo {author}
  {\bibfnamefont {X.-B.}\ \bibnamefont {Wang}}, \bibinfo {author}
  {\bibfnamefont {C.-Z.}\ \bibnamefont {Peng}}, \ and\ \bibinfo {author}
  {\bibfnamefont {J.-W.}\ \bibnamefont {Pan}},\ }\href {\doibase
  10.1038/s41586-020-03093-8} {\bibfield  {journal} {\bibinfo  {journal}
  {Nature}\ }\textbf {\bibinfo {volume} {589}},\ \bibinfo {pages} {214}
  (\bibinfo {year} {2021})}\BibitemShut {NoStop}%
\bibitem [{\citenamefont {Xu}\ \emph {et~al.}(2020)\citenamefont {Xu},
  \citenamefont {Ma}, \citenamefont {Zhang}, \citenamefont {Lo},\ and\
  \citenamefont {Pan}}]{Xu20}%
  \BibitemOpen
  \bibfield  {author} {\bibinfo {author} {\bibfnamefont {F.}~\bibnamefont
  {Xu}}, \bibinfo {author} {\bibfnamefont {X.}~\bibnamefont {Ma}}, \bibinfo
  {author} {\bibfnamefont {Q.}~\bibnamefont {Zhang}}, \bibinfo {author}
  {\bibfnamefont {H.-K.}\ \bibnamefont {Lo}}, \ and\ \bibinfo {author}
  {\bibfnamefont {J.-W.}\ \bibnamefont {Pan}},\ }\href {\doibase
  10.1103/RevModPhys.92.025002} {\bibfield  {journal} {\bibinfo  {journal}
  {Rev. Mod. Phys.}\ }\textbf {\bibinfo {volume} {92}},\ \bibinfo {pages}
  {025002} (\bibinfo {year} {2020})}\BibitemShut {NoStop}%
\bibitem [{\citenamefont {Chen}\ \emph {et~al.}(2020)\citenamefont {Chen},
  \citenamefont {Zhang}, \citenamefont {Liu}, \citenamefont {Jiang},
  \citenamefont {Zhang}, \citenamefont {Hu}, \citenamefont {Guan},
  \citenamefont {Yu}, \citenamefont {Xu}, \citenamefont {Lin}, \citenamefont
  {Li}, \citenamefont {Chen}, \citenamefont {Li}, \citenamefont {You},
  \citenamefont {Wang}, \citenamefont {Wang}, \citenamefont {Zhang},\ and\
  \citenamefont {Pan}}]{Chen20}%
  \BibitemOpen
  \bibfield  {author} {\bibinfo {author} {\bibfnamefont {J.-P.}\ \bibnamefont
  {Chen}}, \bibinfo {author} {\bibfnamefont {C.}~\bibnamefont {Zhang}},
  \bibinfo {author} {\bibfnamefont {Y.}~\bibnamefont {Liu}}, \bibinfo {author}
  {\bibfnamefont {C.}~\bibnamefont {Jiang}}, \bibinfo {author} {\bibfnamefont
  {W.}~\bibnamefont {Zhang}}, \bibinfo {author} {\bibfnamefont {X.-L.}\
  \bibnamefont {Hu}}, \bibinfo {author} {\bibfnamefont {J.-Y.}\ \bibnamefont
  {Guan}}, \bibinfo {author} {\bibfnamefont {Z.-W.}\ \bibnamefont {Yu}},
  \bibinfo {author} {\bibfnamefont {H.}~\bibnamefont {Xu}}, \bibinfo {author}
  {\bibfnamefont {J.}~\bibnamefont {Lin}}, \bibinfo {author} {\bibfnamefont
  {M.-J.}\ \bibnamefont {Li}}, \bibinfo {author} {\bibfnamefont
  {H.}~\bibnamefont {Chen}}, \bibinfo {author} {\bibfnamefont {H.}~\bibnamefont
  {Li}}, \bibinfo {author} {\bibfnamefont {L.}~\bibnamefont {You}}, \bibinfo
  {author} {\bibfnamefont {Z.}~\bibnamefont {Wang}}, \bibinfo {author}
  {\bibfnamefont {X.-B.}\ \bibnamefont {Wang}}, \bibinfo {author}
  {\bibfnamefont {Q.}~\bibnamefont {Zhang}}, \ and\ \bibinfo {author}
  {\bibfnamefont {J.-W.}\ \bibnamefont {Pan}},\ }\href {\doibase
  10.1103/PhysRevLett.124.070501} {\bibfield  {journal} {\bibinfo  {journal}
  {Phys. Rev. Lett.}\ }\textbf {\bibinfo {volume} {124}},\ \bibinfo {pages}
  {070501} (\bibinfo {year} {2020})}\BibitemShut {NoStop}%
\bibitem [{\citenamefont {Briegel}\ \emph {et~al.}(1998)\citenamefont
  {Briegel}, \citenamefont {D\"ur}, \citenamefont {Cirac},\ and\ \citenamefont
  {Zoller}}]{Briegel98}%
  \BibitemOpen
  \bibfield  {author} {\bibinfo {author} {\bibfnamefont {H.-J.}\ \bibnamefont
  {Briegel}}, \bibinfo {author} {\bibfnamefont {W.}~\bibnamefont {D\"ur}},
  \bibinfo {author} {\bibfnamefont {J.~I.}\ \bibnamefont {Cirac}}, \ and\
  \bibinfo {author} {\bibfnamefont {P.}~\bibnamefont {Zoller}},\ }\href
  {\doibase 10.1103/PhysRevLett.81.5932} {\bibfield  {journal} {\bibinfo
  {journal} {Phys. Rev. Lett.}\ }\textbf {\bibinfo {volume} {81}},\ \bibinfo
  {pages} {5932} (\bibinfo {year} {1998})}\BibitemShut {NoStop}%
\bibitem [{\citenamefont {Guha}\ \emph {et~al.}(2015)\citenamefont {Guha},
  \citenamefont {Krovi}, \citenamefont {Fuchs}, \citenamefont {Dutton},
  \citenamefont {Slater}, \citenamefont {Simon},\ and\ \citenamefont
  {Tittel}}]{Guha2015}%
  \BibitemOpen
  \bibfield  {author} {\bibinfo {author} {\bibfnamefont {S.}~\bibnamefont
  {Guha}}, \bibinfo {author} {\bibfnamefont {H.}~\bibnamefont {Krovi}},
  \bibinfo {author} {\bibfnamefont {C.~A.}\ \bibnamefont {Fuchs}}, \bibinfo
  {author} {\bibfnamefont {Z.}~\bibnamefont {Dutton}}, \bibinfo {author}
  {\bibfnamefont {J.~A.}\ \bibnamefont {Slater}}, \bibinfo {author}
  {\bibfnamefont {C.}~\bibnamefont {Simon}}, \ and\ \bibinfo {author}
  {\bibfnamefont {W.}~\bibnamefont {Tittel}},\ }\href {\doibase
  10.1103/PhysRevA.92.022357} {\bibfield  {journal} {\bibinfo  {journal} {Phys.
  Rev. A}\ }\textbf {\bibinfo {volume} {92}},\ \bibinfo {pages} {022357}
  (\bibinfo {year} {2015})}\BibitemShut {NoStop}%
\bibitem [{\citenamefont {Sangouard}\ \emph {et~al.}(2011)\citenamefont
  {Sangouard}, \citenamefont {Simon}, \citenamefont {de~Riedmatten},\ and\
  \citenamefont {Gisin}}]{Sangouard11}%
  \BibitemOpen
  \bibfield  {author} {\bibinfo {author} {\bibfnamefont {N.}~\bibnamefont
  {Sangouard}}, \bibinfo {author} {\bibfnamefont {C.}~\bibnamefont {Simon}},
  \bibinfo {author} {\bibfnamefont {H.}~\bibnamefont {de~Riedmatten}}, \ and\
  \bibinfo {author} {\bibfnamefont {N.}~\bibnamefont {Gisin}},\ }\href
  {\doibase 10.1103/RevModPhys.83.33} {\bibfield  {journal} {\bibinfo
  {journal} {Rev. Mod. Phys.}\ }\textbf {\bibinfo {volume} {83}},\ \bibinfo
  {pages} {33} (\bibinfo {year} {2011})}\BibitemShut {NoStop}%
\bibitem [{\citenamefont {Yu}\ \emph {et~al.}(2020)\citenamefont {Yu},
  \citenamefont {Ma}, \citenamefont {Luo}, \citenamefont {Jing}, \citenamefont
  {Sun}, \citenamefont {Fang}, \citenamefont {Yang}, \citenamefont {Liu},
  \citenamefont {Zheng}, \citenamefont {Xie}, \citenamefont {Zhang},
  \citenamefont {You}, \citenamefont {Wang}, \citenamefont {Chen},
  \citenamefont {Zhang}, \citenamefont {Bao},\ and\ \citenamefont
  {Pan}}]{Yu2020}%
  \BibitemOpen
  \bibfield  {author} {\bibinfo {author} {\bibfnamefont {Y.}~\bibnamefont
  {Yu}}, \bibinfo {author} {\bibfnamefont {F.}~\bibnamefont {Ma}}, \bibinfo
  {author} {\bibfnamefont {X.-Y.}\ \bibnamefont {Luo}}, \bibinfo {author}
  {\bibfnamefont {B.}~\bibnamefont {Jing}}, \bibinfo {author} {\bibfnamefont
  {P.-F.}\ \bibnamefont {Sun}}, \bibinfo {author} {\bibfnamefont {R.-Z.}\
  \bibnamefont {Fang}}, \bibinfo {author} {\bibfnamefont {C.-W.}\ \bibnamefont
  {Yang}}, \bibinfo {author} {\bibfnamefont {H.}~\bibnamefont {Liu}}, \bibinfo
  {author} {\bibfnamefont {M.-Y.}\ \bibnamefont {Zheng}}, \bibinfo {author}
  {\bibfnamefont {X.-P.}\ \bibnamefont {Xie}}, \bibinfo {author} {\bibfnamefont
  {W.-J.}\ \bibnamefont {Zhang}}, \bibinfo {author} {\bibfnamefont {L.-X.}\
  \bibnamefont {You}}, \bibinfo {author} {\bibfnamefont {Z.}~\bibnamefont
  {Wang}}, \bibinfo {author} {\bibfnamefont {T.-Y.}\ \bibnamefont {Chen}},
  \bibinfo {author} {\bibfnamefont {Q.}~\bibnamefont {Zhang}}, \bibinfo
  {author} {\bibfnamefont {X.-H.}\ \bibnamefont {Bao}}, \ and\ \bibinfo
  {author} {\bibfnamefont {J.-W.}\ \bibnamefont {Pan}},\ }\href {\doibase
  10.1038/s41586-020-1976-7} {\bibfield  {journal} {\bibinfo  {journal}
  {Nature}\ }\textbf {\bibinfo {volume} {578}},\ \bibinfo {pages} {240}
  (\bibinfo {year} {2020})}\BibitemShut {NoStop}%
\bibitem [{\citenamefont {Wehner}\ \emph {et~al.}(2018)\citenamefont {Wehner},
  \citenamefont {Elkouss},\ and\ \citenamefont {Hanson}}]{Wehner2018}%
  \BibitemOpen
  \bibfield  {author} {\bibinfo {author} {\bibfnamefont {S.}~\bibnamefont
  {Wehner}}, \bibinfo {author} {\bibfnamefont {D.}~\bibnamefont {Elkouss}}, \
  and\ \bibinfo {author} {\bibfnamefont {R.}~\bibnamefont {Hanson}},\ }\href
  {\doibase 10.1126/science.aam9288} {\bibfield  {journal} {\bibinfo  {journal}
  {Science}\ }\textbf {\bibinfo {volume} {362}},\ \bibinfo {pages} {eaam9288}
  (\bibinfo {year} {2018})}\BibitemShut {NoStop}%
\bibitem [{\citenamefont {Drahi}\ \emph {et~al.}(2021)\citenamefont {Drahi},
  \citenamefont {Sychev}, \citenamefont {Pirov}, \citenamefont {Sazhina},
  \citenamefont {Novikov}, \citenamefont {Walmsley},\ and\ \citenamefont
  {Lvovsky}}]{drahi2021Hybrid}%
  \BibitemOpen
  \bibfield  {author} {\bibinfo {author} {\bibfnamefont {D.}~\bibnamefont
  {Drahi}}, \bibinfo {author} {\bibfnamefont {D.~V.}\ \bibnamefont {Sychev}},
  \bibinfo {author} {\bibfnamefont {K.~K.}\ \bibnamefont {Pirov}}, \bibinfo
  {author} {\bibfnamefont {E.~A.}\ \bibnamefont {Sazhina}}, \bibinfo {author}
  {\bibfnamefont {V.~A.}\ \bibnamefont {Novikov}}, \bibinfo {author}
  {\bibfnamefont {I.~A.}\ \bibnamefont {Walmsley}}, \ and\ \bibinfo {author}
  {\bibfnamefont {A.}~\bibnamefont {Lvovsky}},\ }\href@noop {} {\bibfield
  {journal} {\bibinfo  {journal} {Quantum}\ }\textbf {\bibinfo {volume} {5}},\
  \bibinfo {pages} {416} (\bibinfo {year} {2021})}\BibitemShut {NoStop}%
\bibitem [{\citenamefont {Ulanov}\ \emph {et~al.}(2017)\citenamefont {Ulanov},
  \citenamefont {Sychev}, \citenamefont {Pushkina}, \citenamefont {Fedorov},\
  and\ \citenamefont {Lvovsky}}]{Ulanov2017}%
  \BibitemOpen
  \bibfield  {author} {\bibinfo {author} {\bibfnamefont {A.~E.}\ \bibnamefont
  {Ulanov}}, \bibinfo {author} {\bibfnamefont {D.}~\bibnamefont {Sychev}},
  \bibinfo {author} {\bibfnamefont {A.~A.}\ \bibnamefont {Pushkina}}, \bibinfo
  {author} {\bibfnamefont {I.~A.}\ \bibnamefont {Fedorov}}, \ and\ \bibinfo
  {author} {\bibfnamefont {A.~I.}\ \bibnamefont {Lvovsky}},\ }\href {\doibase
  10.1103/PhysRevLett.118.160501} {\bibfield  {journal} {\bibinfo  {journal}
  {Phys. Rev. Lett.}\ }\textbf {\bibinfo {volume} {118}},\ \bibinfo {pages}
  {160501} (\bibinfo {year} {2017})}\BibitemShut {NoStop}%
\bibitem [{\citenamefont {Sychev}\ \emph {et~al.}(2018)\citenamefont {Sychev},
  \citenamefont {Ulanov}, \citenamefont {Tiunov}, \citenamefont {Pushkina},
  \citenamefont {Kuzhamuratov}, \citenamefont {Novikov},\ and\ \citenamefont
  {Lvovsky}}]{Sychev2018}%
  \BibitemOpen
  \bibfield  {author} {\bibinfo {author} {\bibfnamefont {D.~V.}\ \bibnamefont
  {Sychev}}, \bibinfo {author} {\bibfnamefont {A.~E.}\ \bibnamefont {Ulanov}},
  \bibinfo {author} {\bibfnamefont {E.~S.}\ \bibnamefont {Tiunov}}, \bibinfo
  {author} {\bibfnamefont {A.~A.}\ \bibnamefont {Pushkina}}, \bibinfo {author}
  {\bibfnamefont {A.}~\bibnamefont {Kuzhamuratov}}, \bibinfo {author}
  {\bibfnamefont {V.}~\bibnamefont {Novikov}}, \ and\ \bibinfo {author}
  {\bibfnamefont {A.~I.}\ \bibnamefont {Lvovsky}},\ }\href {\doibase
  10.1038/s41467-018-06055-x} {\bibfield  {journal} {\bibinfo  {journal}
  {Nature Communications}\ }\textbf {\bibinfo {volume} {9}},\ \bibinfo {pages}
  {3672} (\bibinfo {year} {2018})}\BibitemShut {NoStop}%
\bibitem [{\citenamefont {Guccione}\ \emph {et~al.}(2020)\citenamefont
  {Guccione}, \citenamefont {Darras}, \citenamefont {Jeannic}, \citenamefont
  {Verma}, \citenamefont {Nam}, \citenamefont {Cavaillès},\ and\ \citenamefont
  {Laurat}}]{Guccione20}%
  \BibitemOpen
  \bibfield  {author} {\bibinfo {author} {\bibfnamefont {G.}~\bibnamefont
  {Guccione}}, \bibinfo {author} {\bibfnamefont {T.}~\bibnamefont {Darras}},
  \bibinfo {author} {\bibfnamefont {H.~L.}\ \bibnamefont {Jeannic}}, \bibinfo
  {author} {\bibfnamefont {V.~B.}\ \bibnamefont {Verma}}, \bibinfo {author}
  {\bibfnamefont {S.~W.}\ \bibnamefont {Nam}}, \bibinfo {author} {\bibfnamefont
  {A.}~\bibnamefont {Cavaillès}}, \ and\ \bibinfo {author} {\bibfnamefont
  {J.}~\bibnamefont {Laurat}},\ }\href {\doibase 10.1126/sciadv.aba4508}
  {\bibfield  {journal} {\bibinfo  {journal} {Science Advances}\ }\textbf
  {\bibinfo {volume} {6}},\ \bibinfo {pages} {eaba4508} (\bibinfo {year}
  {2020})}\BibitemShut {NoStop}%
\bibitem [{\citenamefont {M\'erolla}\ \emph {et~al.}(1999)\citenamefont
  {M\'erolla}, \citenamefont {Mazurenko}, \citenamefont {Goedgebuer},\ and\
  \citenamefont {Rhodes}}]{PhysRevLett.82.1656}%
  \BibitemOpen
  \bibfield  {author} {\bibinfo {author} {\bibfnamefont {J.-M.}\ \bibnamefont
  {M\'erolla}}, \bibinfo {author} {\bibfnamefont {Y.}~\bibnamefont
  {Mazurenko}}, \bibinfo {author} {\bibfnamefont {J.-P.}\ \bibnamefont
  {Goedgebuer}}, \ and\ \bibinfo {author} {\bibfnamefont {W.~T.}\ \bibnamefont
  {Rhodes}},\ }\href {\doibase 10.1103/PhysRevLett.82.1656} {\bibfield
  {journal} {\bibinfo  {journal} {Phys. Rev. Lett.}\ }\textbf {\bibinfo
  {volume} {82}},\ \bibinfo {pages} {1656} (\bibinfo {year}
  {1999})}\BibitemShut {NoStop}%
\bibitem [{\citenamefont {Gleim}\ \emph {et~al.}(2016)\citenamefont {Gleim},
  \citenamefont {Egorov}, \citenamefont {Nazarov}, \citenamefont {Smirnov},
  \citenamefont {Chistyakov}, \citenamefont {Bannik}, \citenamefont {Anisimov},
  \citenamefont {Kynev}, \citenamefont {Ivanova}, \citenamefont {Collins},
  \citenamefont {Kozlov},\ and\ \citenamefont {Buller}}]{Gleim:16}%
  \BibitemOpen
  \bibfield  {author} {\bibinfo {author} {\bibfnamefont {A.~V.}\ \bibnamefont
  {Gleim}}, \bibinfo {author} {\bibfnamefont {V.~I.}\ \bibnamefont {Egorov}},
  \bibinfo {author} {\bibfnamefont {Y.~V.}\ \bibnamefont {Nazarov}}, \bibinfo
  {author} {\bibfnamefont {S.~V.}\ \bibnamefont {Smirnov}}, \bibinfo {author}
  {\bibfnamefont {V.~V.}\ \bibnamefont {Chistyakov}}, \bibinfo {author}
  {\bibfnamefont {O.~I.}\ \bibnamefont {Bannik}}, \bibinfo {author}
  {\bibfnamefont {A.~A.}\ \bibnamefont {Anisimov}}, \bibinfo {author}
  {\bibfnamefont {S.~M.}\ \bibnamefont {Kynev}}, \bibinfo {author}
  {\bibfnamefont {A.~E.}\ \bibnamefont {Ivanova}}, \bibinfo {author}
  {\bibfnamefont {R.~J.}\ \bibnamefont {Collins}}, \bibinfo {author}
  {\bibfnamefont {S.~A.}\ \bibnamefont {Kozlov}}, \ and\ \bibinfo {author}
  {\bibfnamefont {G.~S.}\ \bibnamefont {Buller}},\ }\href {\doibase
  10.1364/OE.24.002619} {\bibfield  {journal} {\bibinfo  {journal} {Opt.
  Express}\ }\textbf {\bibinfo {volume} {24}},\ \bibinfo {pages} {2619}
  (\bibinfo {year} {2016})}\BibitemShut {NoStop}%
\bibitem [{\citenamefont {Bannik}\ and\ \citenamefont
  {Moiseev}(2021)}]{Bannik:21}%
  \BibitemOpen
  \bibfield  {author} {\bibinfo {author} {\bibfnamefont {O.~I.}\ \bibnamefont
  {Bannik}}\ and\ \bibinfo {author} {\bibfnamefont {E.~S.}\ \bibnamefont
  {Moiseev}},\ }\href {\doibase 10.1364/OE.441619} {\bibfield  {journal}
  {\bibinfo  {journal} {Opt. Express}\ }\textbf {\bibinfo {volume} {29}},\
  \bibinfo {pages} {38858} (\bibinfo {year} {2021})}\BibitemShut {NoStop}%
\bibitem [{\citenamefont {Mel'nik}\ \emph {et~al.}(2018)\citenamefont
  {Mel'nik}, \citenamefont {Arslanov}, \citenamefont {Bannik}, \citenamefont
  {Gilyazov}, \citenamefont {Egorov}, \citenamefont {Gleim},\ and\
  \citenamefont {Moiseev}}]{Melnik2018}%
  \BibitemOpen
  \bibfield  {author} {\bibinfo {author} {\bibfnamefont {K.~S.}\ \bibnamefont
  {Mel'nik}}, \bibinfo {author} {\bibfnamefont {N.~M.}\ \bibnamefont
  {Arslanov}}, \bibinfo {author} {\bibfnamefont {O.~I.}\ \bibnamefont
  {Bannik}}, \bibinfo {author} {\bibfnamefont {L.~R.}\ \bibnamefont
  {Gilyazov}}, \bibinfo {author} {\bibfnamefont {V.~I.}\ \bibnamefont
  {Egorov}}, \bibinfo {author} {\bibfnamefont {A.~V.}\ \bibnamefont {Gleim}}, \
  and\ \bibinfo {author} {\bibfnamefont {S.~A.}\ \bibnamefont {Moiseev}},\
  }\href {\doibase 10.3103/S1062873818080294} {\bibfield  {journal} {\bibinfo
  {journal} {Bulletin of the Russian Academy of Sciences: Physics}\ }\textbf
  {\bibinfo {volume} {82}},\ \bibinfo {pages} {1038} (\bibinfo {year}
  {2018})}\BibitemShut {NoStop}%
\bibitem [{\citenamefont {Samsonov}\ \emph {et~al.}(2020)\citenamefont
  {Samsonov}, \citenamefont {Goncharov}, \citenamefont {Gaidash}, \citenamefont
  {Kozubov}, \citenamefont {Egorov},\ and\ \citenamefont {Gleim}}]{SCW-CV}%
  \BibitemOpen
  \bibfield  {author} {\bibinfo {author} {\bibfnamefont {E.}~\bibnamefont
  {Samsonov}}, \bibinfo {author} {\bibfnamefont {R.}~\bibnamefont {Goncharov}},
  \bibinfo {author} {\bibfnamefont {A.}~\bibnamefont {Gaidash}}, \bibinfo
  {author} {\bibfnamefont {A.}~\bibnamefont {Kozubov}}, \bibinfo {author}
  {\bibfnamefont {V.}~\bibnamefont {Egorov}}, \ and\ \bibinfo {author}
  {\bibfnamefont {A.}~\bibnamefont {Gleim}},\ }\href {\doibase
  10.1038/s41598-020-66948-0} {\bibfield  {journal} {\bibinfo  {journal}
  {Scientific Reports}\ }\textbf {\bibinfo {volume} {10}} (\bibinfo {year}
  {2020}),\ 10.1038/s41598-020-66948-0}\BibitemShut {NoStop}%
\bibitem [{\citenamefont {Chistiakov}\ \emph {et~al.}(2019)\citenamefont
  {Chistiakov}, \citenamefont {Kozubov}, \citenamefont {Gaidash}, \citenamefont
  {Gleim},\ and\ \citenamefont {Miroshnichenko}}]{Chistiakov:19}%
  \BibitemOpen
  \bibfield  {author} {\bibinfo {author} {\bibfnamefont {V.}~\bibnamefont
  {Chistiakov}}, \bibinfo {author} {\bibfnamefont {A.}~\bibnamefont {Kozubov}},
  \bibinfo {author} {\bibfnamefont {A.}~\bibnamefont {Gaidash}}, \bibinfo
  {author} {\bibfnamefont {A.}~\bibnamefont {Gleim}}, \ and\ \bibinfo {author}
  {\bibfnamefont {G.}~\bibnamefont {Miroshnichenko}},\ }\href {\doibase
  10.1364/OE.27.036551} {\bibfield  {journal} {\bibinfo  {journal} {Opt.
  Express}\ }\textbf {\bibinfo {volume} {27}},\ \bibinfo {pages} {36551}
  (\bibinfo {year} {2019})}\BibitemShut {NoStop}%
\bibitem [{\citenamefont {Mora}\ \emph {et~al.}(2012)\citenamefont {Mora},
  \citenamefont {Amaya}, \citenamefont {Ruiz-Alba}, \citenamefont {Martinez},
  \citenamefont {Calvo}, \citenamefont {Mu{\~n}oz},\ and\ \citenamefont
  {Capmany}}]{mora2012simultaneous}%
  \BibitemOpen
  \bibfield  {author} {\bibinfo {author} {\bibfnamefont {J.}~\bibnamefont
  {Mora}}, \bibinfo {author} {\bibfnamefont {W.}~\bibnamefont {Amaya}},
  \bibinfo {author} {\bibfnamefont {A.}~\bibnamefont {Ruiz-Alba}}, \bibinfo
  {author} {\bibfnamefont {A.}~\bibnamefont {Martinez}}, \bibinfo {author}
  {\bibfnamefont {D.}~\bibnamefont {Calvo}}, \bibinfo {author} {\bibfnamefont
  {V.~G.}\ \bibnamefont {Mu{\~n}oz}}, \ and\ \bibinfo {author} {\bibfnamefont
  {J.}~\bibnamefont {Capmany}},\ }\href@noop {} {\bibfield  {journal} {\bibinfo
   {journal} {Optics Express}\ }\textbf {\bibinfo {volume} {20}},\ \bibinfo
  {pages} {16358} (\bibinfo {year} {2012})}\BibitemShut {NoStop}%
\bibitem [{\citenamefont {Bouwmeester}\ \emph {et~al.}(1997)\citenamefont
  {Bouwmeester}, \citenamefont {Pan}, \citenamefont {Mattle}, \citenamefont
  {Eibl}, \citenamefont {Weinfurter},\ and\ \citenamefont
  {Zeilinger}}]{Bouwmeester1997}%
  \BibitemOpen
  \bibfield  {author} {\bibinfo {author} {\bibfnamefont {D.}~\bibnamefont
  {Bouwmeester}}, \bibinfo {author} {\bibfnamefont {J.-W.}\ \bibnamefont
  {Pan}}, \bibinfo {author} {\bibfnamefont {K.}~\bibnamefont {Mattle}},
  \bibinfo {author} {\bibfnamefont {M.}~\bibnamefont {Eibl}}, \bibinfo {author}
  {\bibfnamefont {H.}~\bibnamefont {Weinfurter}}, \ and\ \bibinfo {author}
  {\bibfnamefont {A.}~\bibnamefont {Zeilinger}},\ }\href {\doibase
  10.1038/37539} {\bibfield  {journal} {\bibinfo  {journal} {Nature}\ }\textbf
  {\bibinfo {volume} {390}},\ \bibinfo {pages} {575} (\bibinfo {year}
  {1997})}\BibitemShut {NoStop}%
\bibitem [{\citenamefont {Miroshnichenko}\ \emph {et~al.}(2017)\citenamefont
  {Miroshnichenko}, \citenamefont {Kiselev}, \citenamefont {Trifanov},\ and\
  \citenamefont {Gleim}}]{Miroshnichenko:17}%
  \BibitemOpen
  \bibfield  {author} {\bibinfo {author} {\bibfnamefont {G.~P.}\ \bibnamefont
  {Miroshnichenko}}, \bibinfo {author} {\bibfnamefont {A.~D.}\ \bibnamefont
  {Kiselev}}, \bibinfo {author} {\bibfnamefont {A.~I.}\ \bibnamefont
  {Trifanov}}, \ and\ \bibinfo {author} {\bibfnamefont {A.~V.}\ \bibnamefont
  {Gleim}},\ }\href {\doibase 10.1364/JOSAB.34.001177} {\bibfield  {journal}
  {\bibinfo  {journal} {J. Opt. Soc. Am. B}\ }\textbf {\bibinfo {volume}
  {34}},\ \bibinfo {pages} {1177} (\bibinfo {year} {2017})}\BibitemShut
  {NoStop}%
\bibitem [{\citenamefont {\ifmmode \check{R}\else
  \v{R}\fi{}eh\'a\ifmmode~\check{c}\else \v{c}\fi{}ek}\ \emph
  {et~al.}(2001)\citenamefont {\ifmmode \check{R}\else
  \v{R}\fi{}eh\'a\ifmmode~\check{c}\else \v{c}\fi{}ek}, \citenamefont
  {Hradil},\ and\ \citenamefont {Je\ifmmode~\check{z}\else
  \v{z}\fi{}ek}}]{Rehacek2001}%
  \BibitemOpen
  \bibfield  {author} {\bibinfo {author} {\bibfnamefont {J.}~\bibnamefont
  {\ifmmode \check{R}\else \v{R}\fi{}eh\'a\ifmmode~\check{c}\else
  \v{c}\fi{}ek}}, \bibinfo {author} {\bibfnamefont {Z.}~\bibnamefont {Hradil}},
  \ and\ \bibinfo {author} {\bibfnamefont {M.}~\bibnamefont
  {Je\ifmmode~\check{z}\else \v{z}\fi{}ek}},\ }\href {\doibase
  10.1103/PhysRevA.63.040303} {\bibfield  {journal} {\bibinfo  {journal} {Phys.
  Rev. A}\ }\textbf {\bibinfo {volume} {63}},\ \bibinfo {pages} {040303}
  (\bibinfo {year} {2001})}\BibitemShut {NoStop}%
\bibitem [{\citenamefont {Valivarthi}\ \emph {et~al.}(2020)\citenamefont
  {Valivarthi}, \citenamefont {Davis}, \citenamefont {Pe\~na}, \citenamefont
  {Xie}, \citenamefont {Lauk}, \citenamefont {Narv\'aez}, \citenamefont
  {Allmaras}, \citenamefont {Beyer}, \citenamefont {Gim}, \citenamefont
  {Hussein}, \citenamefont {Iskander}, \citenamefont {Kim}, \citenamefont
  {Korzh}, \citenamefont {Mueller}, \citenamefont {Rominsky}, \citenamefont
  {Shaw}, \citenamefont {Tang}, \citenamefont {Wollman}, \citenamefont {Simon},
  \citenamefont {Spentzouris}, \citenamefont {Oblak}, \citenamefont
  {Sinclair},\ and\ \citenamefont {Spiropulu}}]{PRXQuantum.1.020317}%
  \BibitemOpen
  \bibfield  {author} {\bibinfo {author} {\bibfnamefont {R.}~\bibnamefont
  {Valivarthi}}, \bibinfo {author} {\bibfnamefont {S.~I.}\ \bibnamefont
  {Davis}}, \bibinfo {author} {\bibfnamefont {C.}~\bibnamefont {Pe\~na}},
  \bibinfo {author} {\bibfnamefont {S.}~\bibnamefont {Xie}}, \bibinfo {author}
  {\bibfnamefont {N.}~\bibnamefont {Lauk}}, \bibinfo {author} {\bibfnamefont
  {L.}~\bibnamefont {Narv\'aez}}, \bibinfo {author} {\bibfnamefont {J.~P.}\
  \bibnamefont {Allmaras}}, \bibinfo {author} {\bibfnamefont {A.~D.}\
  \bibnamefont {Beyer}}, \bibinfo {author} {\bibfnamefont {Y.}~\bibnamefont
  {Gim}}, \bibinfo {author} {\bibfnamefont {M.}~\bibnamefont {Hussein}},
  \bibinfo {author} {\bibfnamefont {G.}~\bibnamefont {Iskander}}, \bibinfo
  {author} {\bibfnamefont {H.~L.}\ \bibnamefont {Kim}}, \bibinfo {author}
  {\bibfnamefont {B.}~\bibnamefont {Korzh}}, \bibinfo {author} {\bibfnamefont
  {A.}~\bibnamefont {Mueller}}, \bibinfo {author} {\bibfnamefont
  {M.}~\bibnamefont {Rominsky}}, \bibinfo {author} {\bibfnamefont
  {M.}~\bibnamefont {Shaw}}, \bibinfo {author} {\bibfnamefont {D.}~\bibnamefont
  {Tang}}, \bibinfo {author} {\bibfnamefont {E.~E.}\ \bibnamefont {Wollman}},
  \bibinfo {author} {\bibfnamefont {C.}~\bibnamefont {Simon}}, \bibinfo
  {author} {\bibfnamefont {P.}~\bibnamefont {Spentzouris}}, \bibinfo {author}
  {\bibfnamefont {D.}~\bibnamefont {Oblak}}, \bibinfo {author} {\bibfnamefont
  {N.}~\bibnamefont {Sinclair}}, \ and\ \bibinfo {author} {\bibfnamefont
  {M.}~\bibnamefont {Spiropulu}},\ }\href {\doibase
  10.1103/PRXQuantum.1.020317} {\bibfield  {journal} {\bibinfo  {journal} {PRX
  Quantum}\ }\textbf {\bibinfo {volume} {1}},\ \bibinfo {pages} {020317}
  (\bibinfo {year} {2020})}\BibitemShut {NoStop}%
\bibitem [{\citenamefont {Sibson}\ \emph {et~al.}(2017)\citenamefont {Sibson},
  \citenamefont {Erven}, \citenamefont {Godfrey}, \citenamefont {Miki},
  \citenamefont {Yamashita}, \citenamefont {Fujiwara}, \citenamefont {Sasaki},
  \citenamefont {Terai}, \citenamefont {Tanner}, \citenamefont {Natarajan},
  \citenamefont {Hadfield}, \citenamefont {O'Brien},\ and\ \citenamefont
  {Thompson}}]{Sibson2017}%
  \BibitemOpen
  \bibfield  {author} {\bibinfo {author} {\bibfnamefont {P.}~\bibnamefont
  {Sibson}}, \bibinfo {author} {\bibfnamefont {C.}~\bibnamefont {Erven}},
  \bibinfo {author} {\bibfnamefont {M.}~\bibnamefont {Godfrey}}, \bibinfo
  {author} {\bibfnamefont {S.}~\bibnamefont {Miki}}, \bibinfo {author}
  {\bibfnamefont {T.}~\bibnamefont {Yamashita}}, \bibinfo {author}
  {\bibfnamefont {M.}~\bibnamefont {Fujiwara}}, \bibinfo {author}
  {\bibfnamefont {M.}~\bibnamefont {Sasaki}}, \bibinfo {author} {\bibfnamefont
  {H.}~\bibnamefont {Terai}}, \bibinfo {author} {\bibfnamefont {M.~G.}\
  \bibnamefont {Tanner}}, \bibinfo {author} {\bibfnamefont {C.~M.}\
  \bibnamefont {Natarajan}}, \bibinfo {author} {\bibfnamefont {R.~H.}\
  \bibnamefont {Hadfield}}, \bibinfo {author} {\bibfnamefont {J.~L.}\
  \bibnamefont {O'Brien}}, \ and\ \bibinfo {author} {\bibfnamefont {M.~G.}\
  \bibnamefont {Thompson}},\ }\href {\doibase 10.1038/ncomms13984} {\bibfield
  {journal} {\bibinfo  {journal} {Nature Communications}\ }\textbf {\bibinfo
  {volume} {8}},\ \bibinfo {pages} {13984} (\bibinfo {year}
  {2017})}\BibitemShut {NoStop}%
\bibitem [{\citenamefont {Wang}\ \emph {et~al.}(2018)\citenamefont {Wang},
  \citenamefont {Zhang}, \citenamefont {Chen}, \citenamefont {Bertrand},
  \citenamefont {Shams-Ansari}, \citenamefont {Chandrasekhar}, \citenamefont
  {Winzer},\ and\ \citenamefont {Lon{\v{c}}ar}}]{Wang2018}%
  \BibitemOpen
  \bibfield  {author} {\bibinfo {author} {\bibfnamefont {C.}~\bibnamefont
  {Wang}}, \bibinfo {author} {\bibfnamefont {M.}~\bibnamefont {Zhang}},
  \bibinfo {author} {\bibfnamefont {X.}~\bibnamefont {Chen}}, \bibinfo {author}
  {\bibfnamefont {M.}~\bibnamefont {Bertrand}}, \bibinfo {author}
  {\bibfnamefont {A.}~\bibnamefont {Shams-Ansari}}, \bibinfo {author}
  {\bibfnamefont {S.}~\bibnamefont {Chandrasekhar}}, \bibinfo {author}
  {\bibfnamefont {P.}~\bibnamefont {Winzer}}, \ and\ \bibinfo {author}
  {\bibfnamefont {M.}~\bibnamefont {Lon{\v{c}}ar}},\ }\href {\doibase
  10.1038/s41586-018-0551-y} {\bibfield  {journal} {\bibinfo  {journal}
  {Nature}\ }\textbf {\bibinfo {volume} {562}},\ \bibinfo {pages} {101}
  (\bibinfo {year} {2018})}\BibitemShut {NoStop}%
\bibitem [{\citenamefont {Zhang}\ \emph {et~al.}(2017)\citenamefont {Zhang},
  \citenamefont {Wang}, \citenamefont {Cheng}, \citenamefont {Shams-Ansari},\
  and\ \citenamefont {Lon{\v{c}}ar}}]{zhang2017monolithic}%
  \BibitemOpen
  \bibfield  {author} {\bibinfo {author} {\bibfnamefont {M.}~\bibnamefont
  {Zhang}}, \bibinfo {author} {\bibfnamefont {C.}~\bibnamefont {Wang}},
  \bibinfo {author} {\bibfnamefont {R.}~\bibnamefont {Cheng}}, \bibinfo
  {author} {\bibfnamefont {A.}~\bibnamefont {Shams-Ansari}}, \ and\ \bibinfo
  {author} {\bibfnamefont {M.}~\bibnamefont {Lon{\v{c}}ar}},\ }\href@noop {}
  {\bibfield  {journal} {\bibinfo  {journal} {Optica}\ }\textbf {\bibinfo
  {volume} {4}},\ \bibinfo {pages} {1536} (\bibinfo {year} {2017})}\BibitemShut
  {NoStop}%
\bibitem [{\citenamefont {Boes}\ \emph {et~al.}(2018)\citenamefont {Boes},
  \citenamefont {Corcoran}, \citenamefont {Chang}, \citenamefont {Bowers},\
  and\ \citenamefont {Mitchell}}]{Boes2018}%
  \BibitemOpen
  \bibfield  {author} {\bibinfo {author} {\bibfnamefont {A.}~\bibnamefont
  {Boes}}, \bibinfo {author} {\bibfnamefont {B.}~\bibnamefont {Corcoran}},
  \bibinfo {author} {\bibfnamefont {L.}~\bibnamefont {Chang}}, \bibinfo
  {author} {\bibfnamefont {J.}~\bibnamefont {Bowers}}, \ and\ \bibinfo {author}
  {\bibfnamefont {A.}~\bibnamefont {Mitchell}},\ }\href {\doibase
  https://doi.org/10.1002/lpor.201700256} {\bibfield  {journal} {\bibinfo
  {journal} {Laser \& Photonics Reviews}\ }\textbf {\bibinfo {volume} {12}},\
  \bibinfo {pages} {1700256} (\bibinfo {year} {2018})}\BibitemShut {NoStop}%
\bibitem [{\citenamefont {Miroshnichenko}\ \emph {et~al.}(2018)\citenamefont
  {Miroshnichenko}, \citenamefont {Kozubov}, \citenamefont {Gaidash},
  \citenamefont {Gleim},\ and\ \citenamefont {Horoshko}}]{Miroshnichenko:18}%
  \BibitemOpen
  \bibfield  {author} {\bibinfo {author} {\bibfnamefont {G.~P.}\ \bibnamefont
  {Miroshnichenko}}, \bibinfo {author} {\bibfnamefont {A.~V.}\ \bibnamefont
  {Kozubov}}, \bibinfo {author} {\bibfnamefont {A.~A.}\ \bibnamefont
  {Gaidash}}, \bibinfo {author} {\bibfnamefont {A.~V.}\ \bibnamefont {Gleim}},
  \ and\ \bibinfo {author} {\bibfnamefont {D.~B.}\ \bibnamefont {Horoshko}},\
  }\href {\doibase 10.1364/OE.26.011292} {\bibfield  {journal} {\bibinfo
  {journal} {Opt. Express}\ }\textbf {\bibinfo {volume} {26}},\ \bibinfo
  {pages} {11292} (\bibinfo {year} {2018})}\BibitemShut {NoStop}%
\bibitem [{\citenamefont {Horoshko}\ \emph {et~al.}(2019)\citenamefont
  {Horoshko}, \citenamefont {Eskandari},\ and\ \citenamefont
  {Kilin}}]{HOROSHKO20191728}%
  \BibitemOpen
  \bibfield  {author} {\bibinfo {author} {\bibfnamefont {D.}~\bibnamefont
  {Horoshko}}, \bibinfo {author} {\bibfnamefont {M.}~\bibnamefont {Eskandari}},
  \ and\ \bibinfo {author} {\bibfnamefont {S.}~\bibnamefont {Kilin}},\ }\href
  {\doibase https://doi.org/10.1016/j.physleta.2019.03.006} {\bibfield
  {journal} {\bibinfo  {journal} {Physics Letters A}\ }\textbf {\bibinfo
  {volume} {383}},\ \bibinfo {pages} {1728} (\bibinfo {year}
  {2019})}\BibitemShut {NoStop}%
\bibitem [{\citenamefont {Koashi}(2004)}]{Koashi04}%
  \BibitemOpen
  \bibfield  {author} {\bibinfo {author} {\bibfnamefont {M.}~\bibnamefont
  {Koashi}},\ }\href {\doibase 10.1103/PhysRevLett.93.120501} {\bibfield
  {journal} {\bibinfo  {journal} {Phys. Rev. Lett.}\ }\textbf {\bibinfo
  {volume} {93}},\ \bibinfo {pages} {120501} (\bibinfo {year}
  {2004})}\BibitemShut {NoStop}%
\bibitem [{\citenamefont {Gaidash}\ \emph {et~al.}(2022)\citenamefont
  {Gaidash}, \citenamefont {Miroshnichenko},\ and\ \citenamefont
  {Kozubov}}]{Gaidash:22}%
  \BibitemOpen
  \bibfield  {author} {\bibinfo {author} {\bibfnamefont {A.}~\bibnamefont
  {Gaidash}}, \bibinfo {author} {\bibfnamefont {G.}~\bibnamefont
  {Miroshnichenko}}, \ and\ \bibinfo {author} {\bibfnamefont {A.}~\bibnamefont
  {Kozubov}},\ }\href {\doibase 10.1364/JOSAB.439776} {\bibfield  {journal}
  {\bibinfo  {journal} {J. Opt. Soc. Am. B}\ }\textbf {\bibinfo {volume}
  {39}},\ \bibinfo {pages} {577} (\bibinfo {year} {2022})}\BibitemShut
  {NoStop}%
\bibitem [{\citenamefont {Krovi}\ \emph {et~al.}(2016)\citenamefont {Krovi},
  \citenamefont {Guha}, \citenamefont {Dutton}, \citenamefont {Slater},
  \citenamefont {Simon},\ and\ \citenamefont {Tittel}}]{Krovi2016}%
  \BibitemOpen
  \bibfield  {author} {\bibinfo {author} {\bibfnamefont {H.}~\bibnamefont
  {Krovi}}, \bibinfo {author} {\bibfnamefont {S.}~\bibnamefont {Guha}},
  \bibinfo {author} {\bibfnamefont {Z.}~\bibnamefont {Dutton}}, \bibinfo
  {author} {\bibfnamefont {J.~A.}\ \bibnamefont {Slater}}, \bibinfo {author}
  {\bibfnamefont {C.}~\bibnamefont {Simon}}, \ and\ \bibinfo {author}
  {\bibfnamefont {W.}~\bibnamefont {Tittel}},\ }\href {\doibase
  10.1007/s00340-015-6297-4} {\bibfield  {journal} {\bibinfo  {journal}
  {Applied Physics B}\ }\textbf {\bibinfo {volume} {122}},\ \bibinfo {pages}
  {52} (\bibinfo {year} {2016})}\BibitemShut {NoStop}%
\bibitem [{\citenamefont {Devetak}\ and\ \citenamefont
  {Winter}(2005)}]{Devetak2005}%
  \BibitemOpen
  \bibfield  {author} {\bibinfo {author} {\bibfnamefont {I.}~\bibnamefont
  {Devetak}}\ and\ \bibinfo {author} {\bibfnamefont {A.}~\bibnamefont
  {Winter}},\ }\href {\doibase 10.1098/rspa.2004.1372} {\bibfield  {journal}
  {\bibinfo  {journal} {Proceedings of the Royal Society A: Mathematical,
  Physical and Engineering Sciences}\ }\textbf {\bibinfo {volume} {461}},\
  \bibinfo {pages} {207} (\bibinfo {year} {2005})}\BibitemShut {NoStop}%
\bibitem [{\citenamefont {Assche}(2006)}]{assche_2006}%
  \BibitemOpen
  \bibfield  {author} {\bibinfo {author} {\bibfnamefont {G.~v.}\ \bibnamefont
  {Assche}},\ }\href {\doibase 10.1017/CBO9780511617744} {\emph {\bibinfo
  {title} {Quantum Cryptography and Secret-Key Distillation}}}\ (\bibinfo
  {publisher} {Cambridge University Press},\ \bibinfo {year}
  {2006})\BibitemShut {NoStop}%
\bibitem [{cod()}]{codeSCWDL}%
  \BibitemOpen
  \href@noop {} {\enquote {\bibinfo {title} {See code and details at},}\
  }\bibinfo {howpublished}
  {\url{https://github.com/Eugene91/SCW-DL-Interface}}\BibitemShut {NoStop}%
\bibitem [{\citenamefont {Kiselev}\ \emph {et~al.}(2020)\citenamefont
  {Kiselev}, \citenamefont {Samsonov}, \citenamefont {Goncharov}, \citenamefont
  {Chistyakov}, \citenamefont {Halturinsky}, \citenamefont {Egorov},
  \citenamefont {Kozubov}, \citenamefont {Gaidash},\ and\ \citenamefont
  {Gleim}}]{Kiselev:20}%
  \BibitemOpen
  \bibfield  {author} {\bibinfo {author} {\bibfnamefont {F.}~\bibnamefont
  {Kiselev}}, \bibinfo {author} {\bibfnamefont {E.}~\bibnamefont {Samsonov}},
  \bibinfo {author} {\bibfnamefont {R.}~\bibnamefont {Goncharov}}, \bibinfo
  {author} {\bibfnamefont {V.}~\bibnamefont {Chistyakov}}, \bibinfo {author}
  {\bibfnamefont {A.}~\bibnamefont {Halturinsky}}, \bibinfo {author}
  {\bibfnamefont {V.}~\bibnamefont {Egorov}}, \bibinfo {author} {\bibfnamefont
  {A.}~\bibnamefont {Kozubov}}, \bibinfo {author} {\bibfnamefont
  {A.}~\bibnamefont {Gaidash}}, \ and\ \bibinfo {author} {\bibfnamefont
  {A.}~\bibnamefont {Gleim}},\ }\href {\doibase 10.1364/OE.403293} {\bibfield
  {journal} {\bibinfo  {journal} {Opt. Express}\ }\textbf {\bibinfo {volume}
  {28}},\ \bibinfo {pages} {28696} (\bibinfo {year} {2020})}\BibitemShut
  {NoStop}%
\bibitem [{\citenamefont {Sajeed}\ \emph {et~al.}(2021)\citenamefont {Sajeed},
  \citenamefont {Chaiwongkhot}, \citenamefont {Huang}, \citenamefont {Qin},
  \citenamefont {Egorov}, \citenamefont {Kozubov}, \citenamefont {Gaidash},
  \citenamefont {Chistiakov}, \citenamefont {Vasiliev}, \citenamefont {Gleim},\
  and\ \citenamefont {Makarov}}]{Sajeed2021}%
  \BibitemOpen
  \bibfield  {author} {\bibinfo {author} {\bibfnamefont {S.}~\bibnamefont
  {Sajeed}}, \bibinfo {author} {\bibfnamefont {P.}~\bibnamefont
  {Chaiwongkhot}}, \bibinfo {author} {\bibfnamefont {A.}~\bibnamefont {Huang}},
  \bibinfo {author} {\bibfnamefont {H.}~\bibnamefont {Qin}}, \bibinfo {author}
  {\bibfnamefont {V.}~\bibnamefont {Egorov}}, \bibinfo {author} {\bibfnamefont
  {A.}~\bibnamefont {Kozubov}}, \bibinfo {author} {\bibfnamefont
  {A.}~\bibnamefont {Gaidash}}, \bibinfo {author} {\bibfnamefont
  {V.}~\bibnamefont {Chistiakov}}, \bibinfo {author} {\bibfnamefont
  {A.}~\bibnamefont {Vasiliev}}, \bibinfo {author} {\bibfnamefont
  {A.}~\bibnamefont {Gleim}}, \ and\ \bibinfo {author} {\bibfnamefont
  {V.}~\bibnamefont {Makarov}},\ }\href {\doibase 10.1038/s41598-021-84139-3}
  {\bibfield  {journal} {\bibinfo  {journal} {Scientific Reports}\ }\textbf
  {\bibinfo {volume} {11}},\ \bibinfo {pages} {5110} (\bibinfo {year}
  {2021})}\BibitemShut {NoStop}%
\end{thebibliography}%

\end{document}